\documentclass[fleqn,usenatbib]{mnras}

\usepackage{newtxtext,newtxmath}

\usepackage[T1]{fontenc}

\DeclareRobustCommand{\VAN}[3]{#2}
\let\VANthebibliography\thebibliography
\def\thebibliography{\DeclareRobustCommand{\VAN}[3]{##3}\VANthebibliography}

\usepackage{graphicx}	
\usepackage{amsmath}	
\usepackage[para,online,flushleft]{threeparttable}
\usepackage{multirow}
\usepackage{xcolor}
\usepackage{makecell}
\usepackage{subcaption}
\usepackage{deluxetable}

\newcommand{\mpch}{$h^{-1}$Mpc} 
\newcommand{\kpch}{$h^{-1}$kpc} 

\title[Convergence of small scale structure at $z=5.5$]{Convergence of small scale Ly$\alpha$ structure at high-$z$ under different reionization scenarios}

\author[C.C. Doughty et al.]{
Caitlin C. Doughty$^{1}$\thanks{E-mail: doughty@strw.leidenuniv.nl (CCD)}, 
Joseph F. Hennawi$^{2,1}$, 
Frederick B. Davies$^{3}$, 
Zarija Luki\'c$^{4}$, 
and Jose O\~norbe$^{5}$
\\
$^{1}$Leiden Observatory, Leiden University, P.O. Box 9513, 2300 RA, Leiden, the Netherlands\\
$^{2}$ Department of Physics, University of California, Santa Barbara, CA 93106, USA\\
$^{3}$ Max-Planck-Institut f\"ur Astronomie, K\"onigstuhl 17, D-69117 Heidelberg, Germany \\
$^{4}$ Lawrence Berkeley National Laboratory, CA 94720-8139, USA \\
$^{5}$ Facultad de F\'isicas, Universidad de Sevilla, Avda. Reina Mercedes s/n, Campus Reina Mercedes, E-41012 Seville, Spain
}

\date{Accepted XXX. Received YYY; in original form ZZZ}

\pubyear{2023}

\begin{document}
\label{firstpage}
\pagerange{\pageref{firstpage}--\pageref{lastpage}}
\maketitle

\begin{abstract}
The Ly$\alpha$ forest (LAF) at $z>5$ probes the thermal and reionization history of the intergalactic medium (IGM) and the nature of dark matter, but its interpretation requires comparison to cosmological hydrodynamical simulations. At high-z, convergence of these simulations is more exacting since transmission is dominated by underdense voids that are challenging to resolve. With evidence mounting for a late end to reionization, small structures down to the sub-kpc level may survive to later times than conventionally thought due to the reduced time for pressure smoothing to impact the gas, further tightening simulation resolution requirements. We perform a suite of simulations using the Eulerian cosmological hydrodynamics code \texttt{Nyx}, spanning domain sizes of $1.25-10$ \mpch$\;$ and $5 - 80$ \kpch$\;$ cells, and explore the interaction of these variables with the timing of reionization on the properties of the matter distribution and the simulated LAF at $z=5.5$. In observable Ly$\alpha$ power, convergence within 10\% is achieved for $k<0.1$ s/km, but larger $k$ shows deviation of up to 20 percent. While a later reionization retains more small structure in the density field, because of the greater thermal broadening there is little difference in the convergence of LAF power between early ($z=9$) and later ($z=6$) reionizations. We conclude that at $z\sim5.5$, resolutions of 10 kpc are necessary for convergence of LAF power at $k<0.1$ s/km, while higher-$k$ modes require higher resolution, and that the timing of reionization does not significantly impact convergence given realistic photoheating.
\end{abstract}

\begin{keywords}
keyword1 -- keyword2 -- keyword3
\end{keywords}



\section{Introduction}\label{sec:introduction}
The configuration of matter in the $z>5$ Universe is a probe of many questions of cosmological importance. The initial distribution is set by the primordial power spectrum, describing the density perturbations in space with the overdense regions providing the seeds for later structure formation. Dark matter accounts for the majority of the mass, and so its evolution dominates that of the power spectrum, meaning by extension that the distribution is dependent on the physical properties of dark matter.

While cold dark matter (CDM) is by far the preferred dark matter model due to its success in describing a multitude of observables~\citep[see e.g.][]{bullock17}, its predictions are not always perfectly aligned with observations. Issues such as the missing satellites~\citep{klypin99,moore99}, cusp-core~\citep{flores94,moore94}, and too-big-to-fail~\citep{boylan-kolchin11} problems motivate the exploration of alternative models, such as thermal relic warm dark matter (WDM). Such models, involving lighter particles with masses possibly on the order of a few keV and higher streaming velocities than CDM, would naturally result in a reduction in power on small physical scales. The ability to measure the amplitude of fluctuations on small scales could therefore inform our understanding of the nature of dark matter.

The baryonic matter distribution is initially dominated by that of dark matter, but it is also significantly impacted by major astrophysical events, such as the Epoch of hydrogen Reionization (EoR). The EoR involved the photoionization of the IGM by a buildup of the ultraviolet background (UVB), likely dominated by small galaxies~\citep[though see also][in support of contributions from brighter ones]{naidu20}. Reionization injected an uncertain quantity of heat into the gas, with numerical and theoretical studies estimating the peak temperature to be $\sim$ 20000 K~\citep{miraldaescude94,mcquinn12,daloisio19}. This would have forced a thermodynamic response in the gas and induced an expansion or ``puffing up'' of smaller structure~\citep{katz20,puchwein22}.

The Ly$\alpha$ forest (LAF), measured from quasar spectra, is one available probe of the 1D distribution of matter. Ly$\alpha$ photons emitted by quasars are absorbed by hydrogen gas in the intergalactic medium (IGM) along the line of sight, with the degree of absorption dictated by the gas overdensity, ionization state, temperature, and line-of-sight velocity. The LAF is already well-established as a useful probe of the dark matter distribution and matter power spectrum~\citep{cen94,zhang95,hernquist96,miraldaescude96,theuns98,croft99,viel04,mcdonald06,palanquedelabrouille13,walther18}. The signal is more difficult to extract at smaller physical scales due to resolution limitations, and at higher redshifts both because of the increasing mean density and the increasing neutrality of the gas, which results in substantially reduced transmission and signal-to-noise. Analyses of high resolution, high-S/N spectra have now led to measurements of the Ly$\alpha$ power up to $\log (k/(s$ km$^{-1})) = -1.1$ at $z=5.4$~\citep{viel13b}, and $\log (k/(s$ km$^{-1})) = -0.7$ up to $z=5$~\citep{boera19}.

The interpretation of observations of the LAF power spectrum is complicated by the many unknowns of its sourcing environment. For example, the shape and amplitude of the LAF power spectrum is highly dependent on the IGM temperature at mean density $T_0$ and the UVB; however, $T_0$ is not well constrained at $z>5$~\citep[e.g.][]{walther19}, and investigations of the global H I photoionization rate, for example via the quasar proximity effect~\citep{calverley11} or the mean opacity~\citep{wyithe11,davies18} are themselves contingent on assumption of the temperature. Additionally, the timing and duration of reionization is not certain. While earlier studies established an increase in the Ly$\alpha$ opacity occurring up to $z=6$~\citep{fan06}, often taken to represent the end of the EoR, more recent works have revealed large expanses with little to no Ly$\alpha$ transmission remaining down to $z\sim5.5$~\citep{becker15}. Although a subset of such observations could indicate large scale fluctuations in the temperature or ultraviolet background~\citep{chardin15,chardin17,daloisio15,davies16}, another possibility is that there is still substantial neutral hydrogen present at $z<6$, perhaps with an end to the EoR occurring as late as $z=5.2$~\citep{kulkarni19,keating20}.

In combination, the involved physics require that proper interpretation of LAF observations is necessarily supported by numerical simulations, which can capture nonlinear evolution and account for relevant model variations in gas temperature, UVB, and reionization history. In order to model the Ly$\alpha$ mean flux to full convergence in simulations, it is almost certain that simulation domain sizes of $\sim 40$ \mpch$\;$are necessary~\citep{tytler09,lukic15}. Given this need, and the dynamic range issues intrinsic to simulations, studies of the IGM are typically conducted with an average spatial resolution on the order of 20 \kpch. Small scales are of special importance given that this is where alternative dark matter models will have the largest impact on power, and scales of $\log (k/($km$^{-1}$ s))$\gtrsim -1$ are more sensitive to the process of reionization~\citep{nasir16,onorbe17}.

It is unclear whether 20 \kpch$\;$ is sufficient for proper modeling of the LAF. Reionization and any pre-reionization heating will affect the gas distribution and by extension the LAF by modifying the pressure smoothing scale, $\lambda_\mathrm{ps}$, which arguably must be resolved. This scale is related to the instantaneous comoving Jeans scale, $\lambda_J = \sqrt{\pi c_s^2 / G \rho} \left( 1+z \right)$, where $c_s$ is the sound speed, $G$ the gravitational constant, and $\rho$ the mass density. However, in reality gas cannot react instantaneously to an increase in heat; indeed, the sound crossing time for gas near the mean density is approximately a Hubble time, leading to a sizeable delay in the physical response. Thus, the size at a given time is dependent on the full thermal history of the gas, which can be captured analytically through the filtering scale of~\citet{gnedin98}. Observationally, a filtering or conceptually similar pressure smoothing scale could be inferred using the 1D Ly$\alpha$ flux power spectrum, but it is partially obscured by a degeneracy with the thermal broadening effects present in spectra~\citep{peeples10a,peeples10b,garzilli15}. Methods absent this degeneracy, such as measurements from close quasar pairs, found $\lambda_\mathrm{ps} \approx 90.2_{-30.1}^{+33.4}$ physical kpc at $z=3.6$~\citep{rorai13,rorai17}. At redshifts approaching the end of reionization, this scale will certainly be much smaller.

In this work we explore the convergence of several statistics of structure and the LAF for a suite of cosmological hydrodynamical simulations. In particular, our interest is in determining the resolution and boxsize requirements for reliable interpretation of Ly$\alpha$ observations for large wavenumbers. We also examine how the timing of reionization impacts this convergence. We describe our simulations in Section~\ref{sec:simulations} and results in Section~\ref{sec:results}. We place our findings within the wider context in Section~\ref{sec:discussion}, and provide a summary of the most relevant outcomes in Section~\ref{sec:conclusions}. All reported lengths and length-derived quantities are comoving and include a factor of $h^{-1}$ unless specifically noted.

\section{Simulations}\label{sec:simulations}
For our investigation we use \texttt{Nyx}, a publicly available and highly parallel hydrodynamic cosmological code~\citep{almgren13,sexton21}, built on the adaptive mesh refinement library AMReX~\citep{zhang19}. It uses a Godunov-based piecewise parabolic method~\citep{colella84} to treat the hydrodynamics, which has proved successful in other codes such as {\sc enzo}~\citep{bryan95, bryan14} or {\sc RAMSES}~\citep{teyssier2002}. A geometric multigrid approach is used to solve the Poisson equation for self-gravity of baryons and dark matter, and quantities tracked on the grid are sub-cycled at a rate appropriate for the refinement level. \texttt{Nyx} excludes any treatment of star formation or other galaxy-scale processes, which are generally unnecessary for a proper treatment of the Ly$\alpha$ forest~\citep[see e.g.][]{viel13a}, making it an efficient choice for studies of the IGM. While \texttt{Nyx} includes the capability to incorporate certain feedback processes, such as contributions from AGN, they are not utilized in this study. Neutral and ionized hydrogen, neutral, once-, and twice-ionized helium, and free electrons are tracked, allowing for self-consistent modelling of atomic cooling and UV heating processes.

Another benefit of \texttt{Nyx} is its superior resolution of underdense gas compared to smoothed particle hydrodynamics (SPH) codes. \citet{chabanier22} explored convergence of \texttt{Nyx} against an SPH code {\sc CRK-HACC}~\citep{habib16,frontiere23}. They established that use of the intrinsic SPH kernel in {\sc CRK-HACC} led to biases in flux statistics, in particular as the kernel led to overestimates of the density in underdense regions. They found that the raw Ly$\alpha$ power was more poorly converged in {\sc CRK-HACC} as compared to \texttt{Nyx}, unless they applied a corrective measure to the density field using the Delaunay Tesselation Field Estimator~\citep{schaap07,rangel16}. By using \texttt{Nyx}, we have avoided the complicating factor of a correction to the SPH kernel and ensured more reasonable resolution in the underdense IGM.

\begin{figure}
    \hspace*{-0.2in}
    \includegraphics[width=0.5\textwidth]{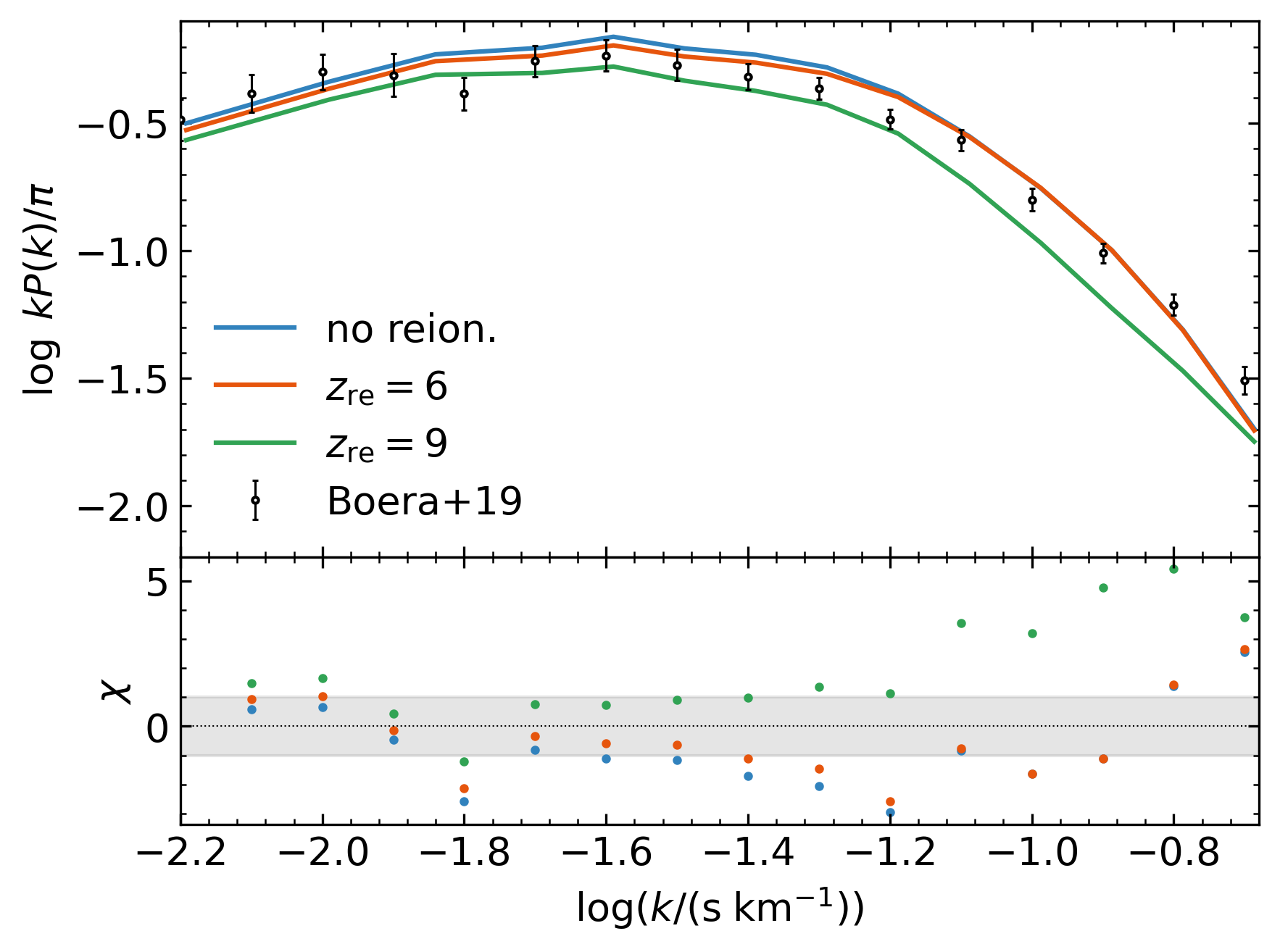}
    \caption{Ly$\alpha$ power at $z=5$ for variable reionization histories, overplotted with $z\sim5$ observations from~\citet{boera19} of the instrumental resolution-corrected power. The residuals in the lower panel are the $\chi$ value between our model predictions and their data points, with the shaded region indicating the 1$\sigma$ spread. The optical depths have been rescaled to match their inferred mean flux value, $\langle F \rangle=\exp \left(- \tau_\mathrm{eff}\right)=0.1845$.}
    \label{fig:boera19_z5_comparison}
\end{figure}

To create the initial conditions, we begin by generating separate spatial and velocity transfer functions in baryons and dark matter using {\sc CAMB}~\citep{lewis00}. The functions are then used as inputs to {\sc CICASS}~\citep{oleary12}, where we presume an initial mean streaming velocity $\langle v \rangle$ between baryons and dark matter of 30 km/s at $z=1000$, approximately the timing of recombination, which decreases to $\sim$6 km/s by $z=200$~\citep{hu95,ma95,tseliakhovich10}. The anticipated low temperature of the pre-reionization IGM, $\sim$ 10 K at $z=100$, could lead to shock heating and entropy generation given that even a sub-1 km/s velocity offset could be supersonic. While many works neglect this velocity offset, given that a main goal of this investigation is to determine the effect of pressure smoothing on Ly$\alpha$ forest convergence, we elect to include it. See Appendix~\ref{sec:appendixb} for a brief analysis of the impact of this choice.

Runtime options in \texttt{Nyx} allow for the implementation of either an inhomogeneous or flash (instantaneous) hydrogen reionization history, with a user-specified maximum quantity of heat $\Delta T_\mathrm{re}$ added at $z_\mathrm{re}$ as in~\citet{onorbe19}
\begin{equation}
    T_f = x_\mathrm{HI} \left( \Delta T_\mathrm{re}-T_i\right) + T_\mathrm{i}
\end{equation}
where $T_f$ and $T_i$ are the final and initial temperatures of a cell, respectively. For example, when an initially fully neutral cell $x_\mathrm{HI}=1$ reionizes, the gas will be heated to a temperature of $\Delta T_\mathrm{re}$. Regions that have already been heated to temperatures higher than $\Delta T_\mathrm{re}$ via shock heating retain these high temperatures. We opt for three flash reionization cases representing (1) an extremely early case at $z_\mathrm{re}=9$, (2) a fiducial case at $z_\mathrm{re}=6$, and (3) an extremely late case, which does not reionize but that does have a temperature floor of $\Delta T_\mathrm{re}$ applied in post-processing. Except where it is noted otherwise, we add 20000 K~\citep{daloisio19}. The details of the simulations analyzed in the main body of this work are noted in Table~\ref{tab:sim_summary}. We assume a $\Lambda$CDM \emph{Planck} cosmology~\citep{planckcol16a}, with $\left(\Omega_M, \Omega_\Lambda, \Omega_b, h, X_H\right)=$ (0.315, 0.685, 0.049, 0.675, 0.751).

\begin{figure}
    \hspace*{0.1in}
    \includegraphics[width=0.5\textwidth]{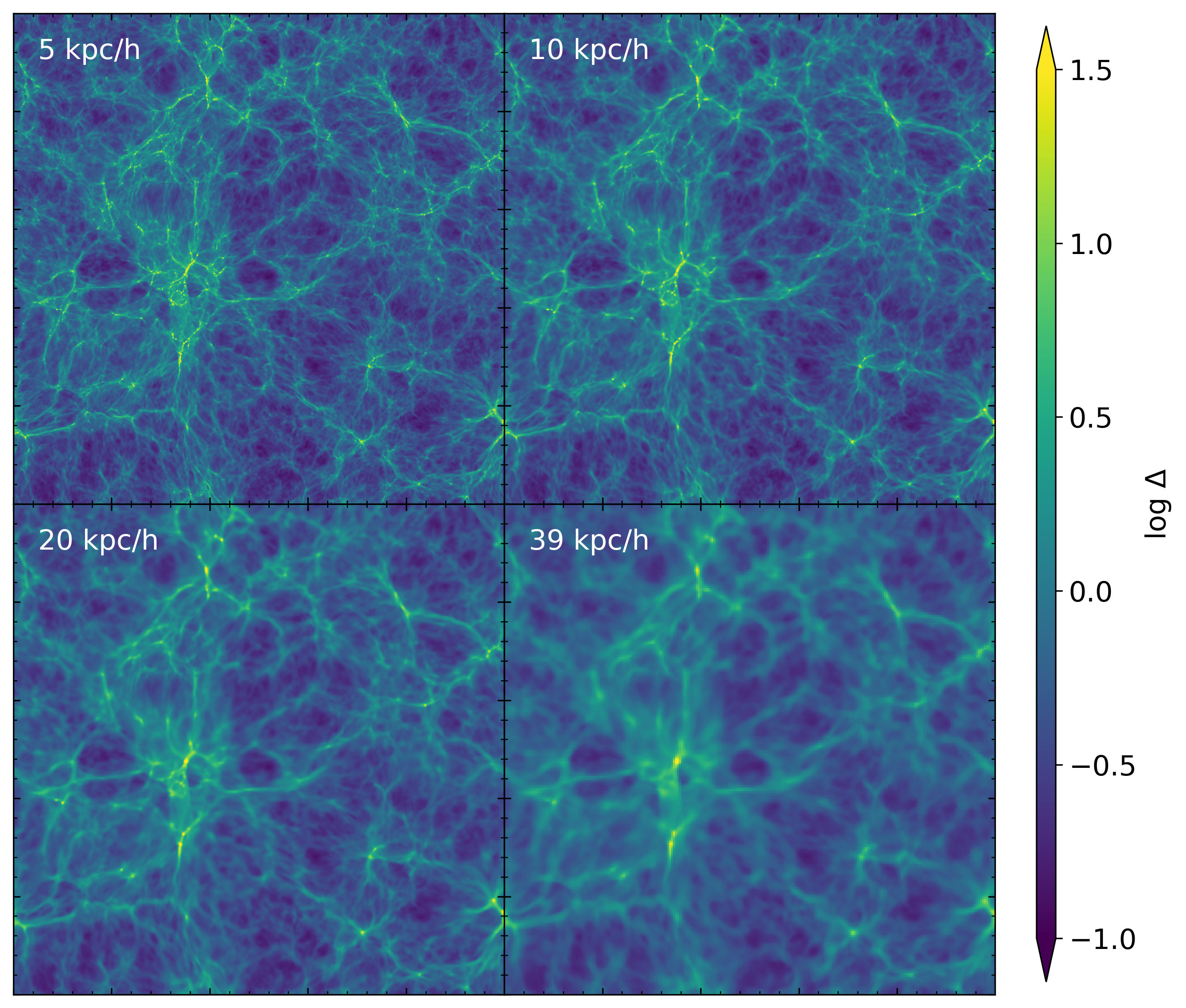}
    \caption{Slices with $\Delta v=$90 km/s of \texttt{Nyx} 10 \mpch$\;$boxes at $z=5.5$ that are flash reionized at $z=6$. Spatial resolutions range from 5 \kpch$\;$at upper left to 40 \kpch$\;$at lower right, with improved resolution resulting in more abundant small scale structure.}
    \label{fig:slices_resolution}
\end{figure}

\begin{table}
  \centering
  \caption{The properties of the simulations used in the main body of this work. The focus of the analysis for a subset of simulations is noted in the leftmost column, and the varied properties within that subset are noted in subsequent columns.}
  \begin{threeparttable}
  \begin{tabular}{ c c c c c }
    \hline
     \makecell{Analysis \\ focus} & \makecell{$L_\mathrm{box}$ \\ (\mpch)} & \makecell{$\Delta x$ \\ (\kpch)} & $z_\mathrm{re}$ & \makecell{$\Delta T_\mathrm{re}$ \\ (K)}\\[0.05in]
     \hline
       \\[-0.08in]
       Resolution & 10 & \makecell{5, 10, \\ 20, 40, 80} & 6 & 20000 \\[0.05in]
       Box size & \makecell{10, 5, \\ 2.5, 1.25} & 5 & 6 & 20000  \\[0.05in]
      \makecell{Reionization \\ history} & 10 & \makecell{5, 10, \\ 20, 40, 80} & 6, 9, none & 5000\tnote{*}, 20000 \\
      \\[-0.08in]
    \hline
  \end{tabular}
  \begin{tablenotes}
  \item[*] An additional $\Delta T_\mathrm{re}$ of 5000 K is only used on the unreionized simulation.
  \end{tablenotes}
  \end{threeparttable}
  \label{tab:sim_summary}
\end{table}

From these simulations we randomly select 10000 lines of sight from the $xy$-plane of the simulation box to serve as the basis for the skewers; the overdensity, temperature, and velocity information are retrieved and used to calculate the opacity in Ly$\alpha$. While \texttt{Nyx} self-consistently tracks the ionization state of hydrogen, we calculate this independently by initially applying a constant hydrogen photoionization rate, $\Gamma_\mathrm{HI}=10^{-12} $s$^{-1}$, to the skewer, and recalculate $x_\mathrm{HI}$ assuming contributions from both photo- and collisional ionization.\footnote{Compare to the actual H I photoheating rate in the model UVB of $3.15 \times 10^{-13}$ s$^{-1}$ at $z=5.5$.} Assuming ionization equilibrium in the IGM may result in differences in the ensuing Ly$\alpha$ power compared to the nonequilibrium case, particularly on scales $\log (k/(\mathrm{s\;km}^{-1}))>-1$~\citep{kusmic22}; however, these differences are largest when reionization is not yet complete and diminish post-reionization.

\begin{figure*}
    \includegraphics[width=1.0\textwidth]{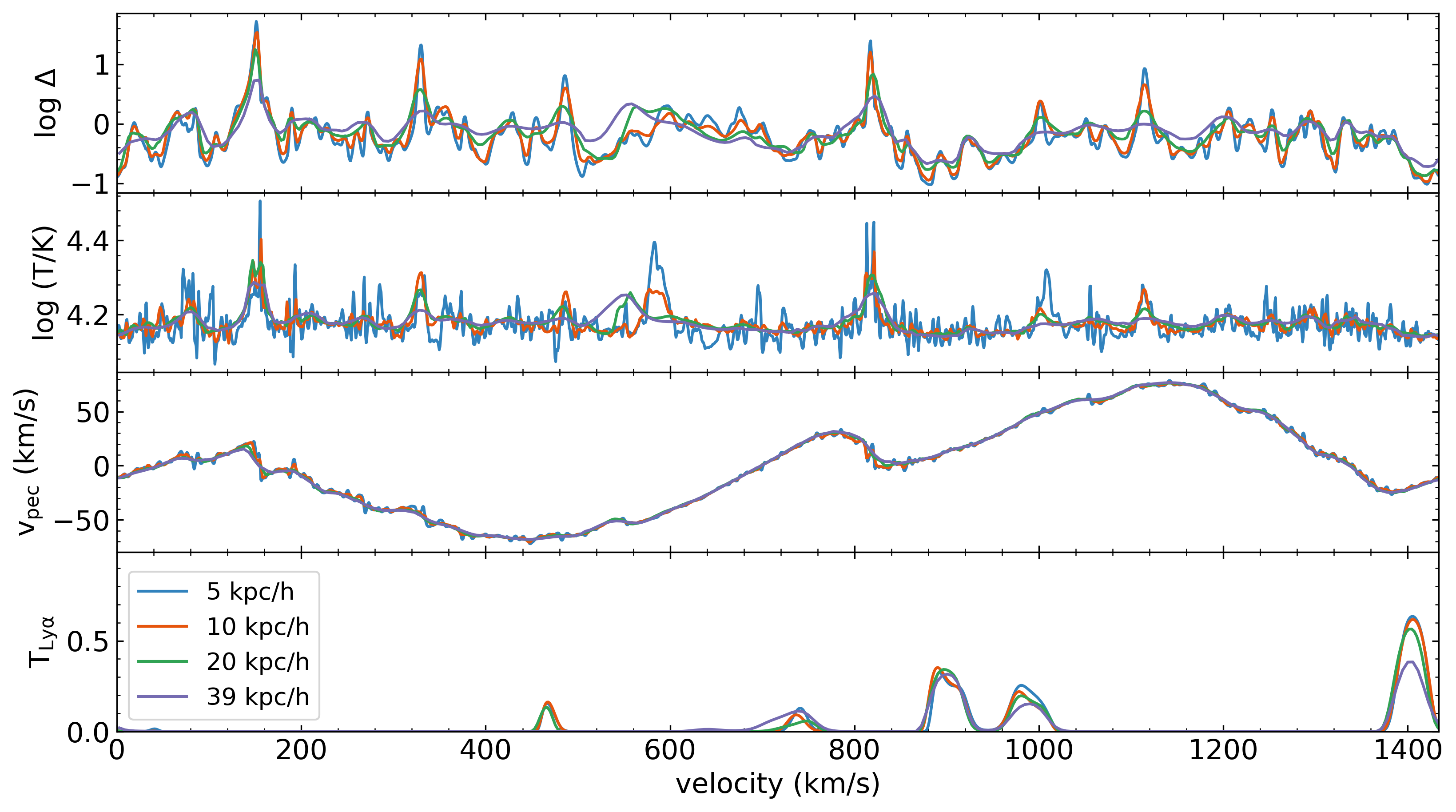}
    \caption{The overdensity, temperature, peculiar velocity, and Ly$\alpha$ transmission fields along a randomly selected skewer in the same simulations as Figure~\ref{fig:slices_resolution}, with varying resolution. Improving spatial resolution from 40 to 5 \kpch$\;$results in more variation with a high spatial frequency, especially visible in overdensity and temperature. These differences cause further variation in the resulting Ly$\alpha$ transmission, in the lowest panel. In particular, excursions to very low overdensities may lead to higher transmission spikes in the bottom panel.}
    \label{fig:skewers_resolution}
\end{figure*}

Ideally, the native spectral resolution of the skewers would be fine enough to well-sample the thermal broadening for the average temperature of the IGM; however, this is not the case for our lower resolution simulations. Thus, we use the method introduced in~\citet{lukic15}, which further discretizes the resolution elements of the native grid while still conserving the total optical depth. In brief, the optical depth of the $j$th spectral pixel as a function of velocity, $\tau_j$, is calculated as the sum of the contributions of the real-space pixels (i.e. the cells)
\begin{equation}\label{eq:tau_calc}
\tau_j = \frac{\pi e^2 f_\mathrm{\mathscr{lu}} \lambda_0}{m_e c H(z)} \sum_i^{N_\mathrm{cells}} \frac{n_{\mathrm{HI},i}}{2} [\mathrm{erf}(y_\mathrm{i-1/2})-\mathrm{erf}(y_\mathrm{i+1/2})]
\end{equation}
where $f_\mathrm{\mathscr{lu}}$ and $\lambda_0$ are the Ly$\alpha$ transition oscillator strength and rest wavelength, $c$ is the speed of light, and $H(z)$ is the Hubble parameter at the relevant redshift. The terms $e$ and $m_e$ are an electron's charge and mass. The total number of real-space cells is given by $N_\mathrm{cells}$, and $y=(v_j-v_{\parallel,\mathrm{pec},i}-v)/b$ describes the shift in line center from the velocity of the pixel in units of the Doppler parameter, $b$. The parallel component of the peculiar velocity is given by $v_{\parallel,\mathrm{pec}}$, and $v$ is the Hubble velocity.

We then perform optical depth rescaling so that the mean flux $\langle F \rangle$ of the skewer set matches an appropriate value given the redshift, which necessitates use of a specific photoionization rate $\Gamma_\mathrm{HI}$. This can be accomplished by generating the spectra for some fiducial $\Gamma_\mathrm{HI, fid}$ and then iteratively re-generating them after adjusting the fiducial value until the mean flux is achieved. Or, if photoionization is assumed to dominate over collisional ionization, one can simply find the factor $A$ such that $\langle F \rangle = \langle \mathrm{exp(-\tau_v/A)} \rangle$ where $\tau_v$ represents the optical depth values acquired using the fiducial photoionization rate. The final $\Gamma_\mathrm{HI}$ is then $A$ times $\Gamma_\mathrm{HI,fid}$.

In Figure~\ref{fig:boera19_z5_comparison} we compare Ly$\alpha$ power at $z=5$ for our three simplified reionization scenarios to measurements from~\citet{boera19} in simulations with $L_\mathrm{box}=10$ \mpch$\;$and spatial resolution $\Delta x=5$ \kpch, with the optical depth rescaled to match their inferred $\tau_\mathrm{eff}=1.69$. We plot $\chi$ in the lower panel, with $\chi=(\log P_\mathrm{data}-\log P_\mathrm{model})/\sigma_\mathrm{data}$. The models are all within about $\pm 2 \chi$ of the data for $\log(k/($s km$^{-1}))<-1.3$,  with the exception of the low $\log (k/($s km$^{-1}))=-1.8$ bin. For larger $k$, the $z_\mathrm{re}=6$ and \textit{no reion.} models are a better match to the observations.

\section{Results}\label{sec:results}
In this section we examine the effects of resolution, box size, and reionization history on various simulation metrics related to the configuration of matter and the Ly$\alpha$ forest. In subsections~\ref{ssec:resolution_effects} and \ref{ssec:boxsize_effects}, we focus on the effects of resolution and box size, respectively, used in \texttt{Nyx} in the case of a fiducial reionization history, $z_\mathrm{re}=6$. In subsection~\ref{ssec:reionization_history} we explore how the specific implementation of reionization affects the requirements to achieve convergence in the pertinent metrics.

\subsection{Resolution effects when $z_\mathrm{re}=6$}\label{ssec:resolution_effects}
We begin by visualizing the impact of resolution with a fiducial reionization history, specifically one in which reionization occurred late so that small IGM structures have not had much time to dissipate by $z=5.5$. Compared to an earlier reionization history, this could in theory make the convergence requirements more stringent. Figure~\ref{fig:slices_resolution} shows 90 km/s slices from a 10 \mpch$\;$box with spatial resolutions varying from 5 to 40 \kpch, and Figure~\ref{fig:skewers_resolution} shows the physical quantities along a sample skewer from the same simulations. Modeling a higher spatial resolution allows for the formation of smaller structures in both the underlying dark matter and in the gas that gives rise to Ly$\alpha$ absorption. These structures are visible by eye in the slices as sharper features, and in the physical quantities along the skewer as narrower fluctuations in overdensity, temperature, velocity, and the emergent quantity of Ly$\alpha$ transmission, $T_\mathrm{Ly\alpha}$.

\begin{figure}
\hspace*{-0.1in}
\includegraphics[width=0.5\textwidth]{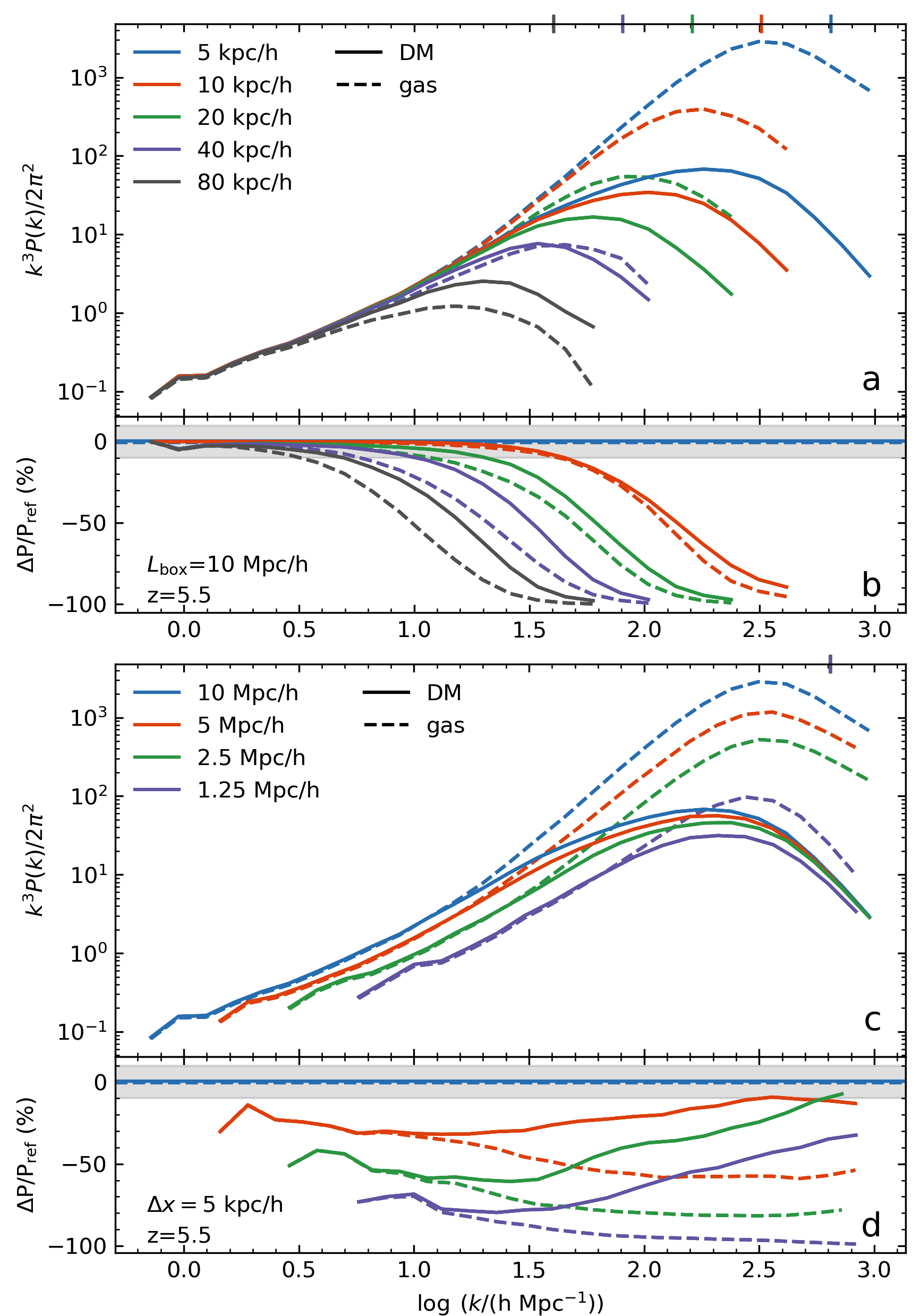}
\caption{\emph{Top panel:} The 3D matter density for baryons (dashed) and dark matter (solid) in a 10 \mpch$\;$box with $z_\mathrm{re}=9$ and varying spatial resolution. On the panel sharing the x-axis, we show the residuals for each baryon and dark matter 3D power curve with respect to the corresponding curve from the highest resolution simulation, here $\sim$5 \kpch. The Nyquist sampling frequency, dictated by the spatial resolution of the box, is indicated with the ticks on the upper x axis. \emph{Bottom panel:} The same as in upper panel, but for varying box sizes. Here, residuals are shown with respect to the largest box, $L_\mathrm{box}=10$ \mpch.} 
\label{fig:P3d_resolution_boxsize}
\end{figure}

To visualize the effect of resolution on the density field, we plot the 3D matter power spectra at $z=5.5$ in gas and dark matter as a function of the wavenumber (panel \textit{a} of Figure~\ref{fig:P3d_resolution_boxsize}). Both the gas and dark matter power increase as the spatial resolution is improved, i.e. as $\Delta x$ decreases. Additionally, improved resolution causes the power in gas to surpass that of dark matter in the same simulation at small scales, with the crossover point occurring at decreasing values of $\log k$ as $\Delta x$ is decreased. For 80 \kpch, the gas power is always lower than the dark matter power, but by $\Delta x= 40$ \kpch$\;$there is a turnover occurring at $\log (k/(h$ Mpc$^{-1}))=1.6$ above which gas power is higher. This abundance of small structure is not likely related to the Ly$\alpha$ forest, however; it arises from high density structures. This is in part due to the absence of galaxy formation models in \texttt{Nyx}, as high density regions which would be forming galaxies and producing stellar feedback are instead permitted to cool and collapse without interruption~\citep[see for example Figure 2 and associated discussion in][]{chabanier22}. These could be removed with an overdensity clip~\citep[see][]{kulkarni15}, inducing a similar effect to {\sc quicklya} in MP-Gadget~\citep{feng18}, but we do not apply this, and instead simply make an out-of-the-box comparison.

\begin{figure}
\hspace*{-0.1in}
\includegraphics[width=0.5\textwidth]{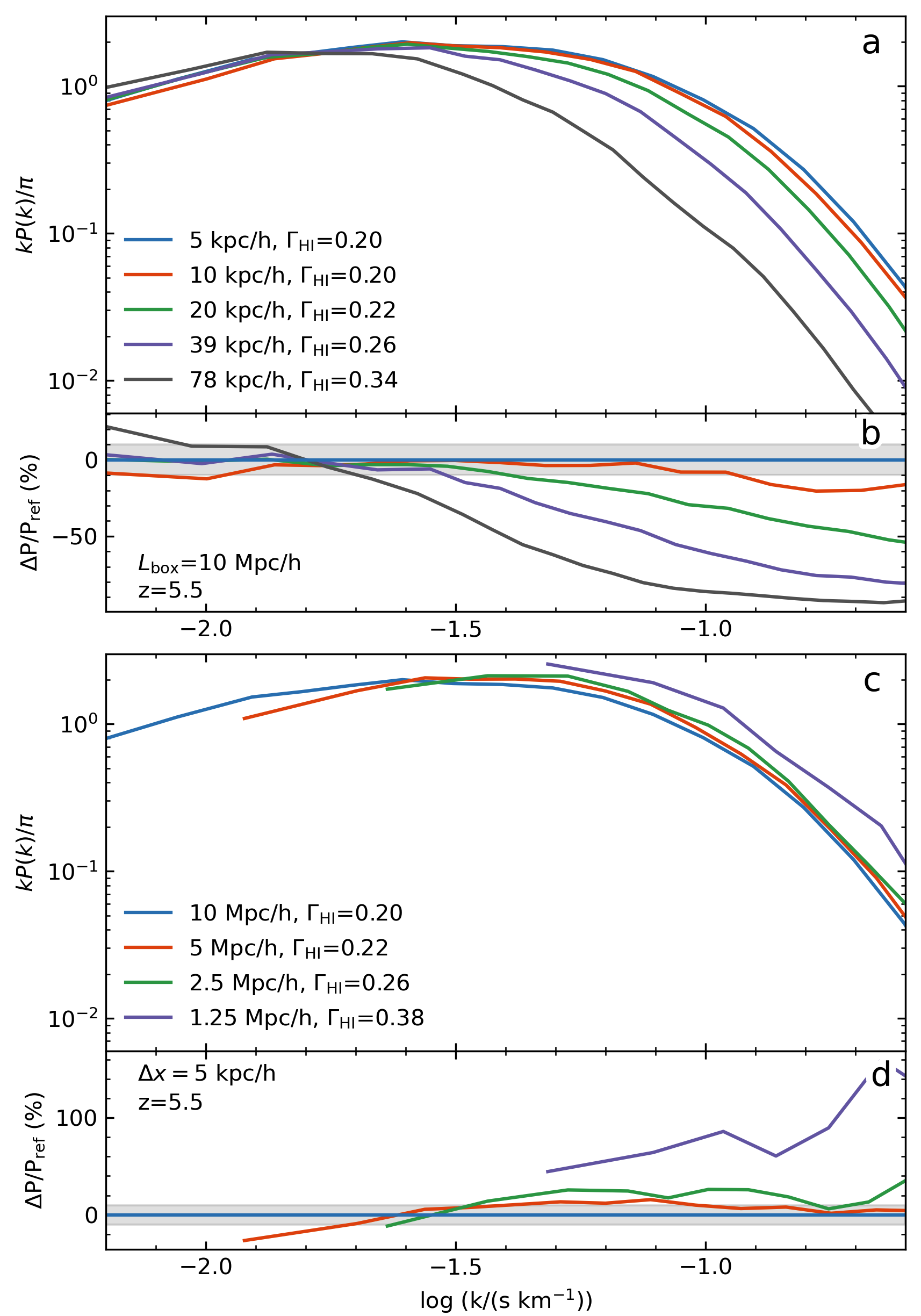}
\caption{\emph{Top panel:} The 1D Ly$\alpha$ forest power spectra for skewers taken from the 10 \mpch$\;$box with $z_\mathrm{re}=6$ and varying spatial resolution. The $\Gamma_\mathrm{HI}$ values in the legend indicate the photoionization rate necessary to rescale the optical depths to the observed average at $z=5.5$ of $\langle F \rangle$=0.05~\citep{eilers18,bosman21}. Immediately below are the residuals of each power spectrum with respect to the highest resolution curve. \emph{Bottom panel:} For box sizes $2.5 < L_\mathrm{box} < 10$ \mpch$\;$with $\Delta x=5$ \kpch, at the largest scales the smaller boxes show lower rescaled power compared to the large box, likely resulting from both an undersampling of large modes and the lack of large-scale mode coupling. The smallest $L=1.25$ \mpch$\;$box sits significantly above the other rescaled power curves. \emph{Note:} The smallest $k_\mathrm{Nyq}$ for these models is $\log (k/$(s km$^{-1}))=-0.55$, which is just off the right edge of the plot.}
\label{fig:PLya_resolution_boxsize}
\end{figure}

The convergence trend is illustrated in panel \textit{b} of Figure~\ref{fig:P3d_resolution_boxsize}, showing the residuals between each DM or baryon power curve with respect to the equivalent 5 \kpch$\;$resolution simulation. The 10 \kpch$\;$ simulation converges within 10 per cent for $\log (k/(h$ Mpc$^{-1}))<1.6$, corresponding to a comoving size of $\sim$0.23 \mpch, but the lowest resolution simulations are lacking a substantial amount of power. For example, at $\log (k/(h$ Mpc$^{-1}))=1.5$, the gas in the 80 \kpch$\;$resolution box has only a few per cent of the power in $\Delta x=5$ \kpch. The simulation with 20 \kpch$\;$resolution, commonly used in IGM studies, is still about 30 per cent low in gas density power at this scale compared to 5 \kpch.

In a Universe with increased underlying small scale structure in the dark matter and gas, it is expected that some of this additional structure should be visible in the Ly$\alpha$ forest power spectrum, as long as variability is enhanced in the appropriate overdensities, $-1 \lesssim \log \Delta \lesssim 0$ for $z\sim5$~\citep{chabanier22}. Accordingly, we plot the 1D Ly$\alpha$ forest power spectra in Figure~\ref{fig:PLya_resolution_boxsize}, calculated from 10000 \texttt{Nyx} skewers as discussed in Section~\ref{sec:simulations}. Here, the optical depths are rescaled to reproduce $\langle F \rangle=0.05$, appropriate for $z=5.5$~\citep{bosman21}, and the 1D dimensionless power in Ly$\alpha$ is calculated from the flux contrast field $\delta = F/\langle F \rangle - 1$.

\begin{figure*}
\includegraphics[width=0.8\textwidth]{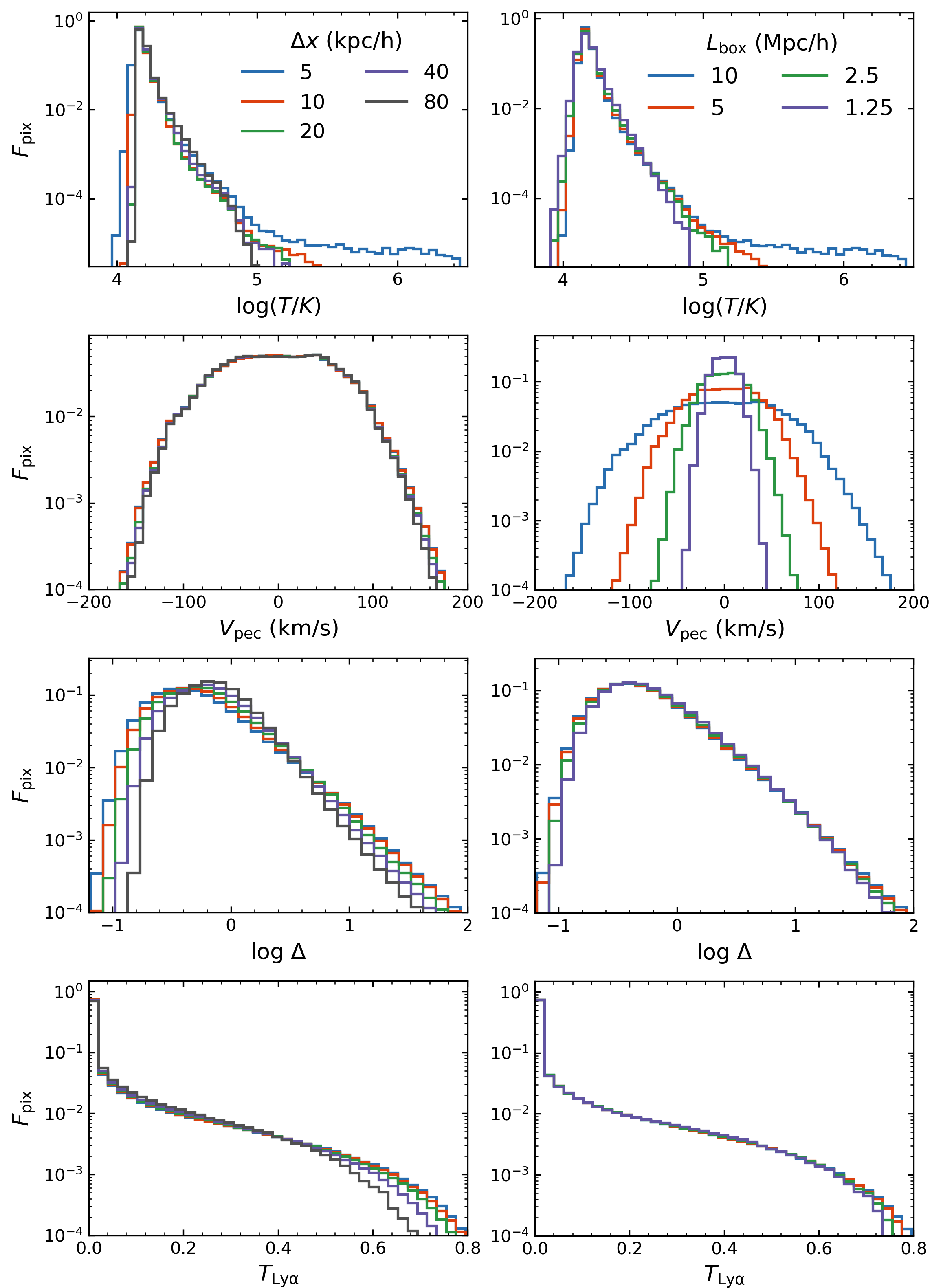}
\caption{\emph{Left column:} The fraction of pixels $F_\mathrm{pix}$ along 10000 skewers as a function of the temperature, peculiar velocity, overdensity, and mean flux-adjusted transmission for varying spatial resolutions. Generally, the higher resolution simulations show wider ranges in their pixel physical properties. \emph{Right column:} Same as the left column, but for varying box sizes with a constant spatial resolution of 5 \kpch. Larger boxes result in a larger number of excursions to more extreme values in temperature, velocity, overdensity, and flux, similar to the trend seen with increasing resolution.}
\label{fig:pdf_resolution_boxsize}
\end{figure*}

\begin{figure*}
\centering
\includegraphics[width=1.0\textwidth]{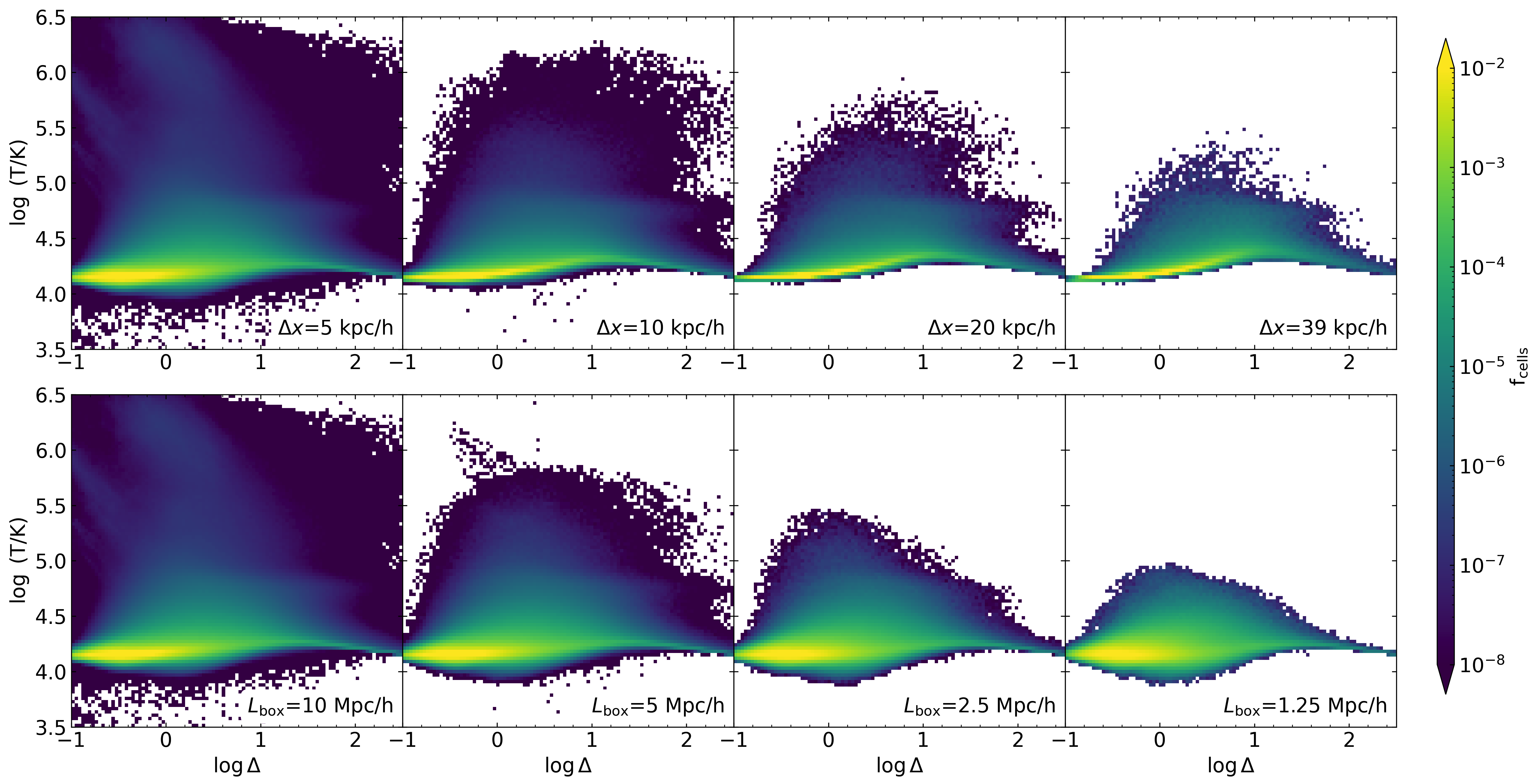}
\caption{\emph{Top row:} Phase diagrams for 10 \mpch$\;$boxes with flash reionization at $z_\mathrm{re}=6$, with the fraction of cells $f_\mathrm{cells}$ indicated by color. Better spatial resolutions result in a wider range of possible physical values for the temperature and density, in particular at higher temperatures, for a fixed range of $f_\mathrm{cells}$. However, these represent a small fraction of the total cells. \emph{Bottom row:} Same as in top row, but for varying box sizes with constant resolution of 5 \kpch$\;$(the upper left and lower left panels are identical). Moving to smaller box sizes results in fewer excursions to extreme values, especially visible in temperature. Generally for variation in resolution and box size, a reduction in the number of cells used in the model causes a more restricted range of gas properties to be sampled.}
\label{fig:rhot_resolution_boxsize}
\end{figure*}

We find that the large scale power $\log (k/(\mathrm{s\;km}^{-1}))<-1.7$ is converged for $\Delta x<40$ \kpch, but that 80 \kpch$\;$has excessive power. At smaller scales, as seen with the 3D matter density field, the power increases with improved spatial resolution. It is not quite converged by 10 \kpch, with the residual between 10 and 5 \kpch$\;$being $\sim$ 15 per cent for $\log (k/($s km$^{-1}))>-1.0$. We also note the $\Gamma_\mathrm{HI}$ rates that are required to match the mean flux; these are noted in the legend in Figure~\ref{fig:PLya_resolution_boxsize} in units of $10^{-12}$ s$^{-1}$. We see that as the spatial resolution increases, the photoionization rate required to match the observed mean flux decreases; for example, from 80 to 20 \kpch$\;$resolution $\Gamma_\mathrm{HI}$ drops by a factor of 1.7. Thus, for a given photoionization rate, the high resolution simulations naturally produce more transmission compared to the low resolution versions. 

Using the values along the skewers, the normalized probability distribution functions in the temperature, peculiar velocity, overdensity, and flux (with optical depth rescaling) are shown in the left column of Figure~\ref{fig:pdf_resolution_boxsize} for different $\Delta x$. The main result is that lower resolution simulations show an excess $F_\mathrm{pix}$ at intermediate values in temperature and overdensity with respect to the 5 \kpch$\;$resolution model. As the resolution improves, these excesses diminish. For the overdensity the excess peaks near the mean density $\log \Delta=0$, highlighting the tendency of a lower resolution simulation to sample more ``average'' conditions due to larger cell sizes. However, in temperature this excess occurs at $\log (T$/K)$\sim4.5$, which does not encompass the average temperature in these boxes, so some differences may be due to variation in the physics permitted by the differing spatial resolutions. For example, compared even to 10 \kpch, there is a tail of extreme high temperatures in the 5 \kpch$\;$simulation that seems to be due to an increased ability to resolve shocks. Generally though, the higher resolution permits excursions to more extreme physical conditions.

The broader physical spread permitted by the higher resolution simulation is visible also in phase diagrams in the top row of Figure~\ref{fig:rhot_resolution_boxsize}. The typical shape is the same since reionization occurred quite recently in this model; the temperature-density relation is nearly flat as the average temperature is approximately 15000 K (though dependent on a cell's neutral fraction). While the 10 \kpch$\;$simulation contains cells reaching $\log (T$/K)$>6$, the 40 \kpch$\;$run can barely achieve temperatures of $\log (T$/K)$>5.0$. This is a combination of the effects of physics due to the different resolutions, but also due to the total number of cells.

The wider variance in physical properties in the high resolution simulation lends itself to the distribution of transmissions shown in the bottom panel of Figure~\ref{fig:pdf_resolution_boxsize}. With increasing resolution, a larger fraction of pixels have either zero flux or $T_\mathrm{Ly\alpha}>0.3$. The higher number of high transmission pixels are created by the higher fraction of $\log \Delta < 0$ cells, and these also contribute to the lower photoionization rate required to match the mean flux. Conversely, the greater abundance of zero flux pixels arises from the higher number of overdense cells in the high resolution simulations, but their abundance does not significantly affect the mean flux because of their high optical depths. The greater variance about the mean flux also ultimately contributes to the greater power seen in the higher resolution simulations, since $\sigma^2$ is proportional to the integral over the power spectrum.

\subsection{Box size effects when $z_\mathrm{re}=6$}\label{ssec:boxsize_effects}
The 3D power in matter with varying simulation box size is plotted in panel c of Figure~\ref{fig:P3d_resolution_boxsize}, for a constant $\Delta x=5$ \kpch. For each box size, the dark matter and gas curves for each box size converge together for $\log (k/(h$ Mpc$^{-1}))<0.8$, but for larger $\log k$ the gas generally deviates more, in all cases exceeding $P_\mathrm{DM}$. Similar to the trend with decreasing $\Delta x$, we see that a larger box size affects the relationship between the dark matter and gas, where the power in gas will exceed that in dark matter at increasingly small $k$ for larger boxes. The power is not yet converged within 10 per cent by $L_\mathrm{box}=10$ \mpch$\;$for any physical scale considered here, and still differs by $\sim25$ per cent between $L_\mathrm{box}=5$ and 10 \mpch.

\begin{figure}
\includegraphics[width=0.5\textwidth]{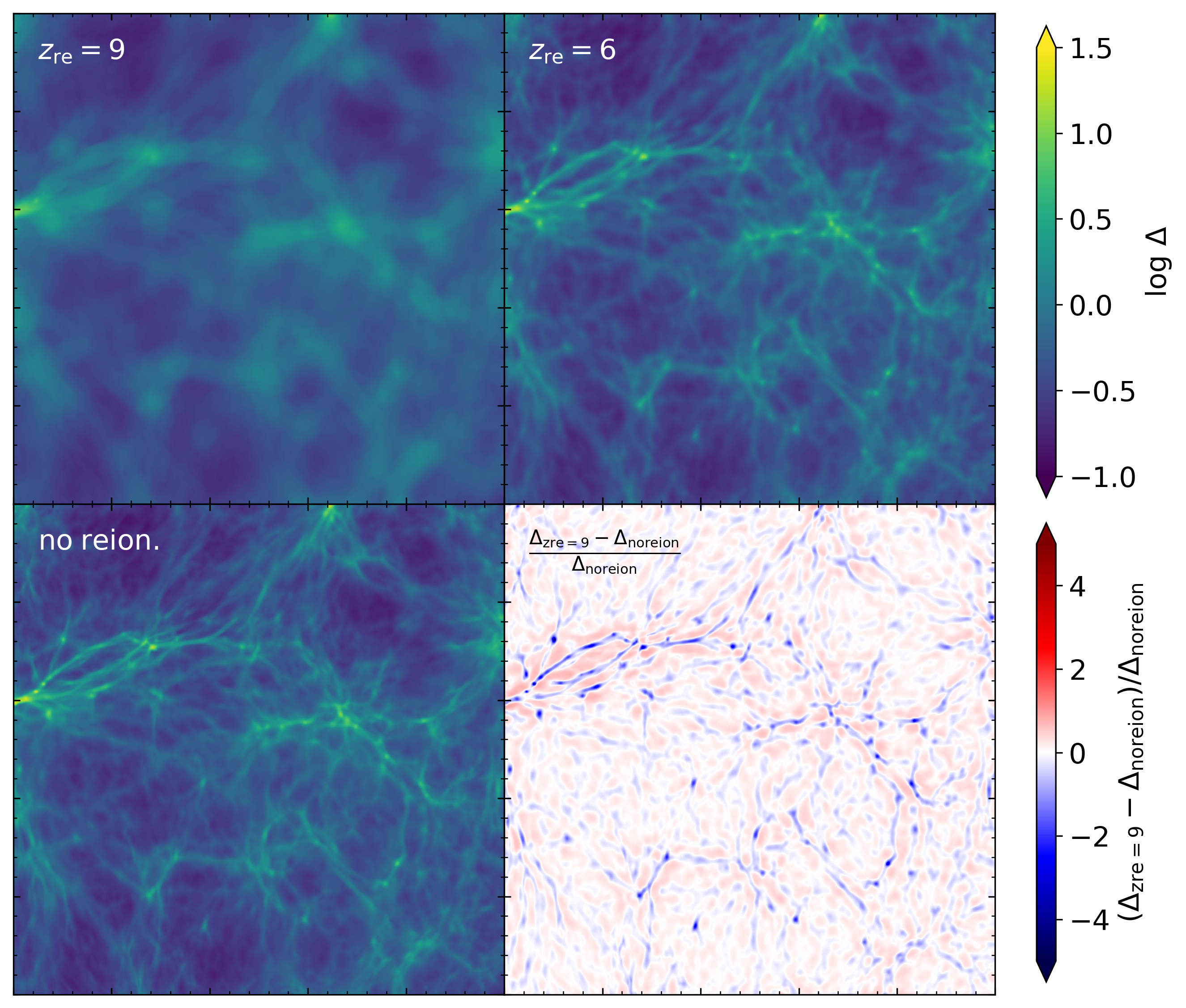}
\caption{Zoomed-in 2.5 \mpch$\;$overdensity slices with $\Delta v=90$ km/s, taken from 10 \mpch$\;$boxes with spatial resolution of $\sim$5 \kpch$\;$and varying reionization histories. Two are flash reionized, at $z=9$ and $=6$, and one is not reionized but has a temperature floor of 20000 K applied at $z=5.5$ in post-processing to mimic a very late reionization. The $z_\mathrm{re}=9$ box has lower temperatures and higher transmission than the later reionization histories. The lower right panel shows the ratio of the $z_\mathrm{re}=9$ box with respect to the \emph{no reion.} simulation.}
\label{fig:slices_reion_hist}
\end{figure}

Examining the Ly$\alpha$ power convergence for these same models in the lower panels of Figure~\ref{fig:PLya_resolution_boxsize}, we find that the smaller the box the greater the normalization of the power curve, such that e.g. the 1.25 \mpch$\;$box has 80 per cent more power than the 10 \mpch$\;$box at $\log (k/$(s km$^{-1}))=-1.0$. By $L_\mathrm{box}=5$ \mpch, the power is converged within 10 per cent for $\log (k/$(s km $^{-1}$))$\leq-1.7$, with some random variation around the 0 per cent residual line. It is possible then that studies of small scale structure may be permissible in small boxes where the larger scale power is not yet converged, although the rate of convergence is not necessarily clear by $L_\mathrm{box}=10$ \mpch. The $\Gamma_\mathrm{HI}$ values, noted in the legend of Figure~\ref{fig:PLya_resolution_boxsize}, required to match $\langle F \rangle$ decrease with increasing box size, by a factor of 1.5 from $L_\mathrm{box}=1.25$ to 10 \mpch. In the right column of Figure~\ref{fig:pdf_resolution_boxsize}, the overdensity distributions show a similar trend to that seen in the PDFs' variation with resolution, where a larger box (like for a higher resolution) shows that a wider range of overdensities are simulated. This trend also proves true for high temperatures, but not for low ones, with smaller boxes showing an excess in $F_\mathrm{pix}$ with $\log (T/K)<4.2$ with respect to the 10 \mpch$\;$box.

\subsection{Effects of differing reionization histories on simulation convergence}\label{ssec:reionization_history}
We now turn to examine the changes induced by varying the reionization history. In total, we consider two flash implementations occurring at $z_\mathrm{re}=9$ and 6 representing an early and late (fiducial) reionization, and \emph{no reion.} indicating a box which is not reionized but that has $\Delta T_\mathrm{re}=20000$ K added in post-processing prior to calculation of the Ly$\alpha$ transmission for the skewers. To provide some visual intuition for the effects, Figure~\ref{fig:slices_reion_hist} displays slices in overdensity from three 10 \mpch$\;$boxes with 5 \kpch$\;$resolution that have varying reionization implementations. The more recent the reionization event, the less time permitted for the IGM to thermodynamically respond to the reionization heating event. Thus, our selected histories represent cases with ever smaller pressure smoothing lengths, and thus more stringent convergence requirements.

\begin{figure*}
\centering
\includegraphics[width=1.0\textwidth]{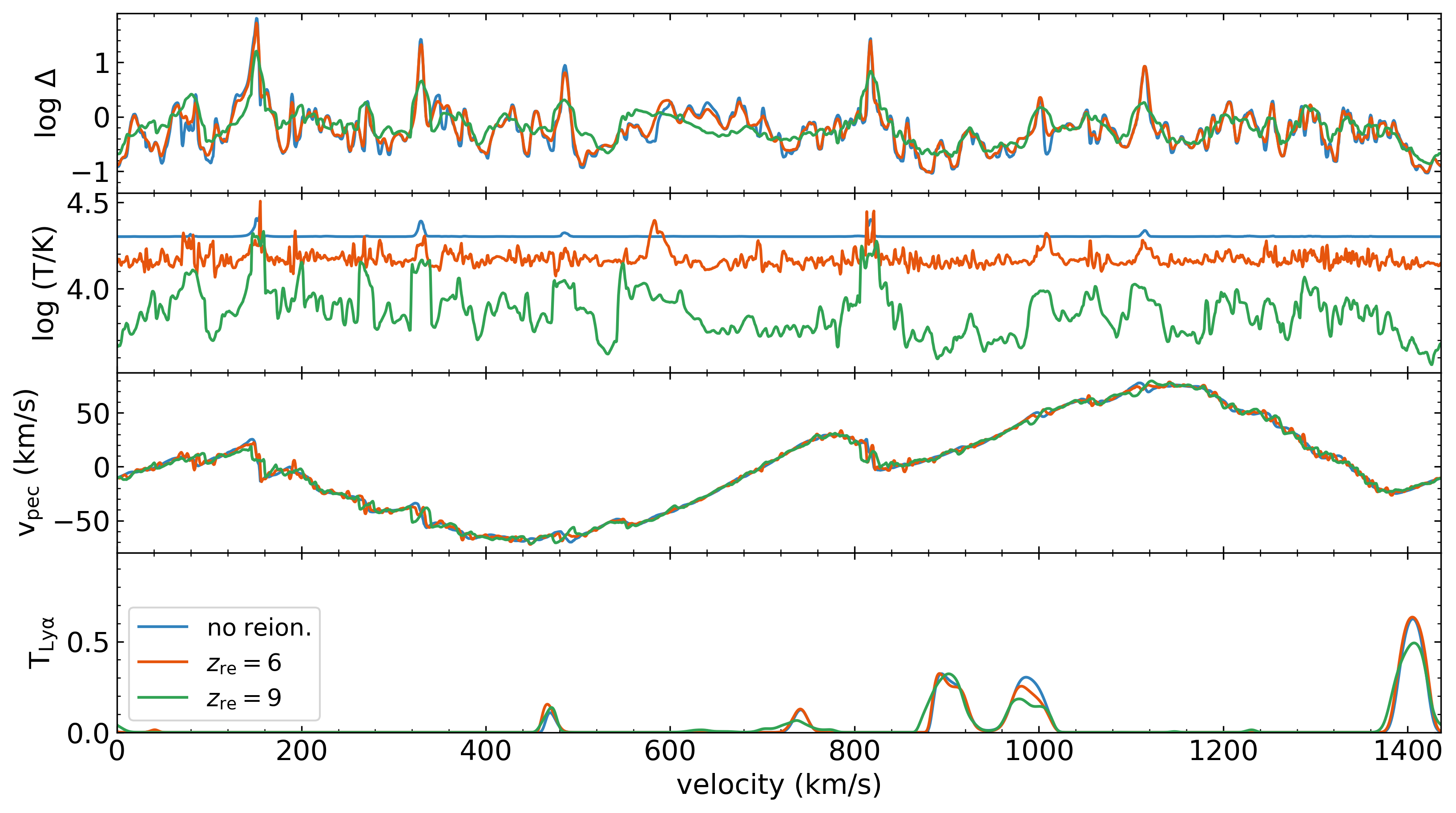}
\caption{Skewers selected from 10 \mpch$\;$boxes with 5 \kpch$\;$resolution with our three simple reionization histories. All models have the same amount of heat added at reionization, $T_\mathrm{re}=20000$ K, and so for the \emph{no reion.} model especially this leads to a nearly isothermal skewer at the injection temperature. Comparing the extremes, \emph{no reion.} and $z_\mathrm{re}=9$, many narrow variations in density present in the former are extended in the latter, for example in the $\log \Delta$ field from 800-1000 km/s. In the optical depth-rescaled $T_\mathrm{Ly\alpha}$ skewer in the bottom panel, this translates into higher, narrower peaks in transmission in the later reionizations.}
\label{fig:skewers_reion_hist}
\end{figure*}

Examining the difference in the overdensity fields between the $z_\mathrm{re}=9$ and \emph{no reion.} cases (lower right panel of Figure~\ref{fig:slices_reion_hist}), we see that there are significant losses of material from the central regions of filaments with an earlier flash, and gains on their exteriors which are spatially more extended but smaller in magnitude. The increasing pressure smoothing scale is clearly visible in the slice panels of Figure~\ref{fig:slices_reion_hist} as a ``blurring'' effect that increases as the redshift of reionization increases. In the skewer panels in Figure~\ref{fig:skewers_reion_hist} this same effect is visible as a smoothing of the variation in the overdensity, temperature, and peculiar velocity progressing from later to earlier flash reionization. The broadening effect is also visible in the Ly$\alpha$ transmission, with the $z_\mathrm{re}=9$ showing wider transmission spikes that are also reduced in amplitude.

\begin{figure*}
\includegraphics[width=1.0\textwidth]{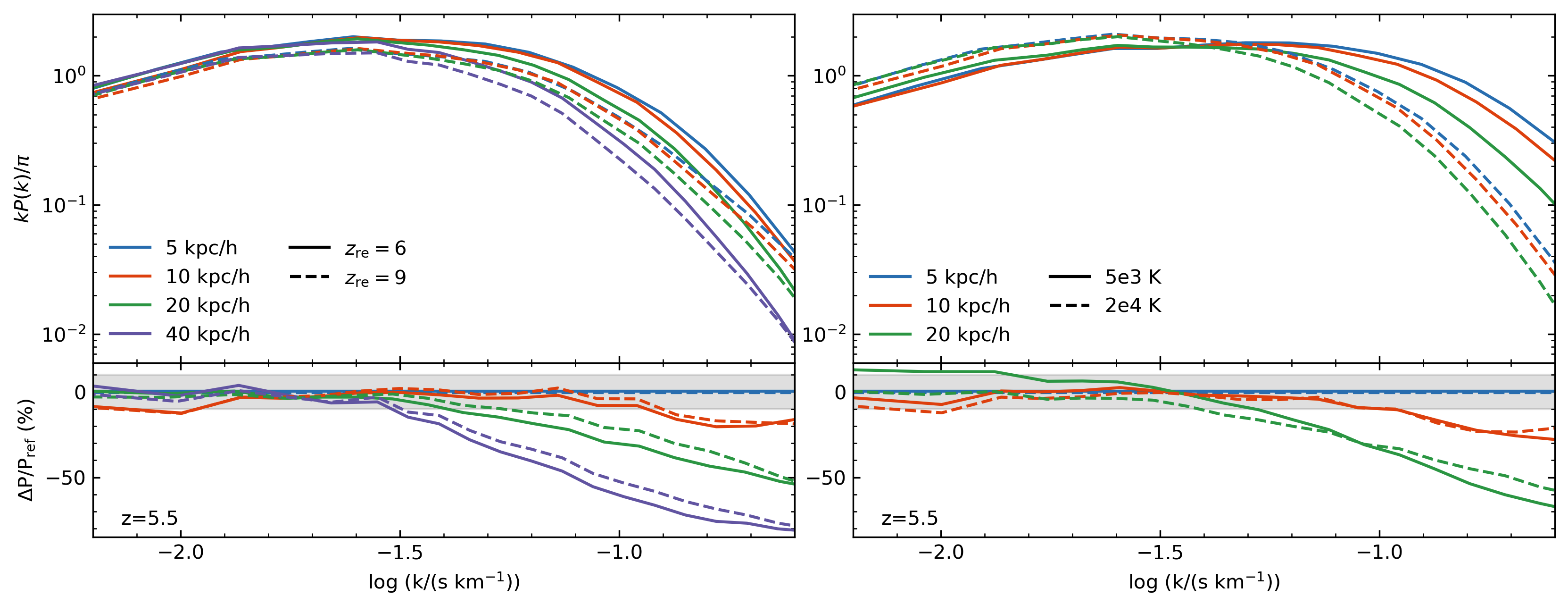}
\caption{Ly$\alpha$ power spectra in 10 \mpch$\;$boxes with varying reionization histories and changes with resolution. The left panel shows the two flash reionizations (with the $z_\mathrm{re}=6$ curves being identical to those in Figure~\ref{fig:PLya_resolution_boxsize}), while the right panel shows the curves for the \emph{no reion.} simulation with two different amounts of heat injected in post-processing. Residuals in the bottom panels are taken with respect to the 5 \kpch$\;$models, and the 5 \kpch$\;$curves are shown with a slight offset from zero for visibility. The shaded gray region indicates the $\pm$ 10 per cent region.}
\label{fig:PLya_reion_hist}
\end{figure*}

The Ly$\alpha$ power spectra and residuals for varying resolutions in the two flash reionization histories are shown in the left panel of Figure~\ref{fig:PLya_reion_hist}; for each reionization and resolution combination, the residuals in the lower panel are shown with respect to the $\Delta x=5$ \kpch$\;$case for each reionization model. First, the $P_\mathrm{Ly\alpha}$ for both flash reionization scenarios increases with resolution for $\log (k/$(s km$^{-1}$))$>-1.5$ with $P_\mathrm{Ly\alpha}(\Delta x=40$ \kpch) lying more than 60 per cent below the $\Delta x=5$ \kpch$\;$curve for $\log (k/$s km$^{-1}$)$>-1.0$. The models all converge within 10 per cent for $\log (k/$(s km$^{-1}$))$<-1.6$. Despite the differences seen in the 3D power in Figure~\ref{fig:pressure_smoothing}, the rate of convergence is comparable for these two histories, and the residuals for each with respect to the highest resolution simulation do not differ from one another by more than about 5 per cent. The residuals are consistently smallest for $z_\mathrm{re}=9$, showing a \emph{slightly} faster convergence.

As seen in the skewers in Figure~\ref{fig:skewers_reion_hist}, there are substantial differences in the temperatures of the $z_\mathrm{re}=6$ and 9 models; on average, the later flash is more than twice as hot, which naturally will lead to greater thermal broadening. To explore the impact of temperature, we plot $P_\mathrm{Ly\alpha}$ for the \emph{no reion.} simulation with both 5000 K and the fiducial 20000 K of heat painted on in the right panel of Figure~\ref{fig:PLya_reion_hist}. For 20 \kpch$\;$resolution and $\log (k/(\mathrm{s\;km}^{-1})>-1.0$, the 20000 K skewers lie more than 25 per cent below the 5000 K ones, though by 10 \kpch$\;$this diminishes by about half. However, as with the two flash scenarios, the residual differences between the two heating models with respect to the 5 \kpch$\;$simulations do not differ all that much from each other, only by $\sim10$ per cent for 20 \kpch$\;$and $5$ per cent for 10 \kpch, showing a similar rate of convergence.

Given that the \emph{no reion.} case visibly contains the most small scale fluctuations in overdensity in Figures~\ref{fig:slices_reion_hist} and~\ref{fig:skewers_reion_hist}, it is worth remarking that Figure~\ref{fig:PLya_reion_hist} shows higher small scale Ly$\alpha$ power in the $z_\mathrm{re}=6$ case than in \emph{no reion.} with $\Delta T_\mathrm{re}=20000$ K, lying a few per cent higher. One would intuitively expect to see the effects of a larger pressure smoothing scale in the $z_\mathrm{re}=6$ reionization scenario manifest as diminishing power at larger $\log k$ compared to the unreionized box.

To demonstrate explicitly that it is primarily the temperature causing this reversal in the expected trend, we explore the impact of varying the other relevant physical parameters. First, we take the skewer set from the \emph{no reion.} simulation (with the reduced heat injection of 5000 K) and re-calculate $P_\mathrm{Ly\alpha}$; this is the reference case, intended to be affected minimally by thermal broadening. Next, we take these same skewers and replace the peculiar velocity, overdensity, and both of these values on all skewers with those from the two flash reionization scenarios, and calculate the power spectra ($P_\mathrm{V}$, $P_\mathrm{\Delta}$, $P_\mathrm{V,\Delta}$). \footnote{It should be noted that this is not physically realistic, as the replacement does not preserve the appropriate post-reionization temperature-density relation, or any other physical relationship between these quantities. It is only an illustrative example to isolate the contribution of each physical parameter to the power spectrum residual.}

\begin{figure*}
\includegraphics[width=0.8\textwidth]{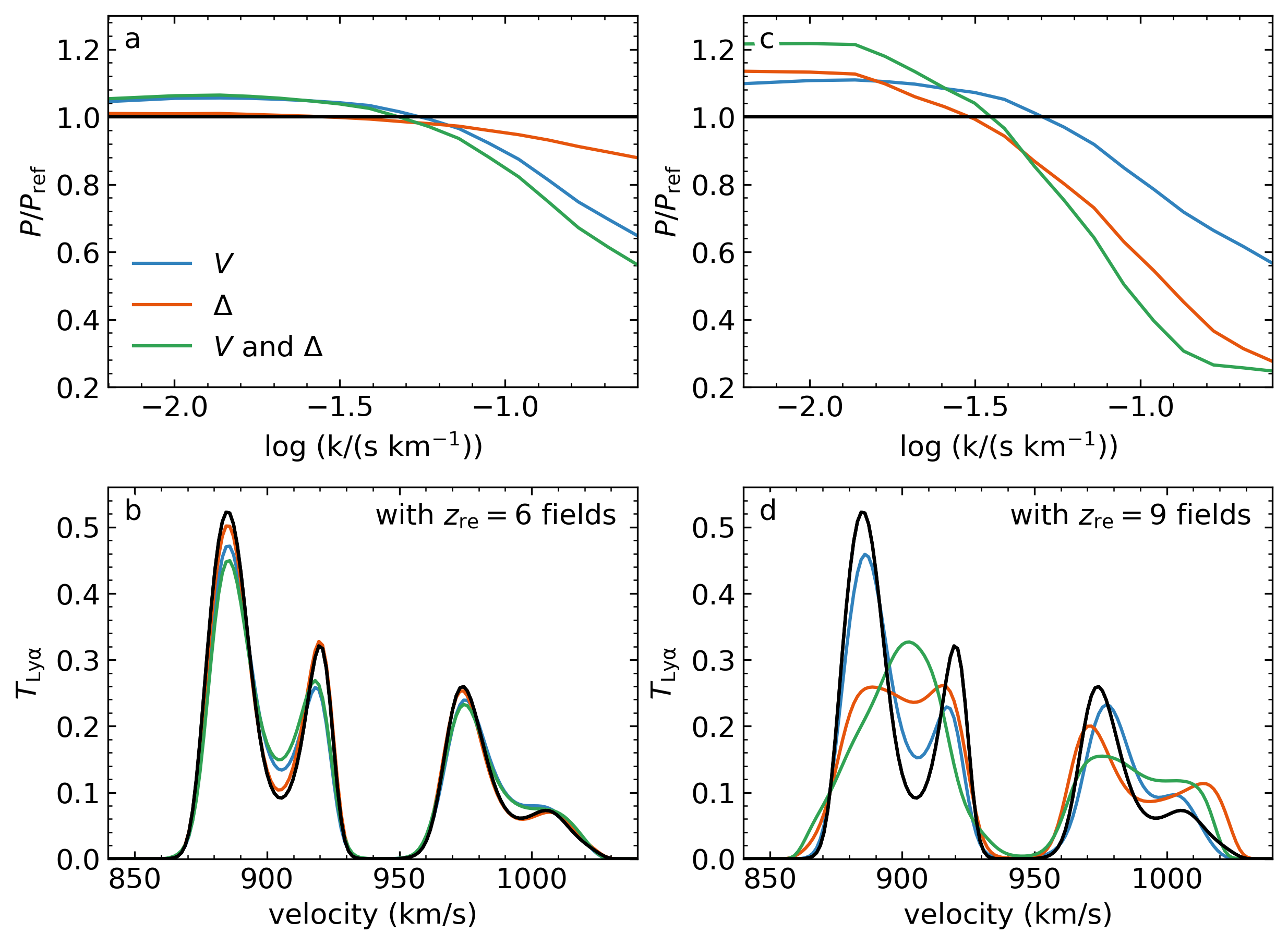}
\caption{\emph{Top row:} Ly$\alpha$ power spectra residuals for the unreionized skewers when replacing fields with values taken from a flashed simulation. Replacement fields are taken from the $z_\mathrm{re}=6$ (left panel) and $z_\mathrm{re}=9$ (right panel) simulations, and the replaced fields are indicated by color. \emph{Bottom panels:} $T_\mathrm{Ly\alpha}$ on a segment of an example skewer with the modified fields in colored lines, and the reference unreionized skewer shown in black. Note that these panels are zoomed-in for improved visibility, and do not reach $T_\mathrm{Ly\alpha}=1$.}
\label{fig:residuals_with_replacement}
\end{figure*}

In Figure~\ref{fig:residuals_with_replacement}, we show the ratios of the raw and modified power spectra, along with the variation of an example skewer segment in response to the replacements. With respect to our $\sim$ 5000 K, \emph{no reion.} reference skewer, each replacement considered here results in a reduction in small scale power at $\log k\gtrapprox-1.4$ and gains at larger scales. When replacing with quantities from the $z_\mathrm{re}=6$ skewers (left column), the power (panel a) decreases progressively with $P_\Delta$, $P_\mathrm{V}$, and $P_\mathrm{V,\Delta}$. On the other hand, replacing the overdensity field with that from the earlier flash simulation causes a larger drop in power than replacing the velocity field (panel c), though of course replacing both fields still results in a greater loss of power. We can see the reason for the reversal in the impact of replacing $\Delta$ and $V$ in the example skewers in Figure~\ref{fig:skewers_reion_hist}: In the top panel the overdensities in $z_\mathrm{re}=6$ are clearly very similar to those in \emph{no reion.}, with the only visible deviations in $T_\mathrm{Ly\alpha}$ visible occurring at the non-zero local minima and maxima along the curve. In $z_\mathrm{re}=9$, however, the overdensity field is altered, leading to broadened transmission features, obviously a byproduct of the expanded spatial distribution of H I in that reionization scenario. From these results, it is apparent that the peculiar velocities and overdensities of the earlier reionization models both alone and in combination cause a loss in small scale power. Therefore, the increase in small scale power of the $z_\mathrm{re}=6$ simulation with respect to \emph{no reion}. with $\Delta T_\mathrm{re}=20000$ K seen in Figure~\ref{fig:PLya_reion_hist} arises solely from the confounding effect of temperature.

\subsection{Measuring the pressure smoothing scale}\label{ssec:measuring_pss}
\begin{figure*}
\centering
\includegraphics[width=1.0\textwidth]{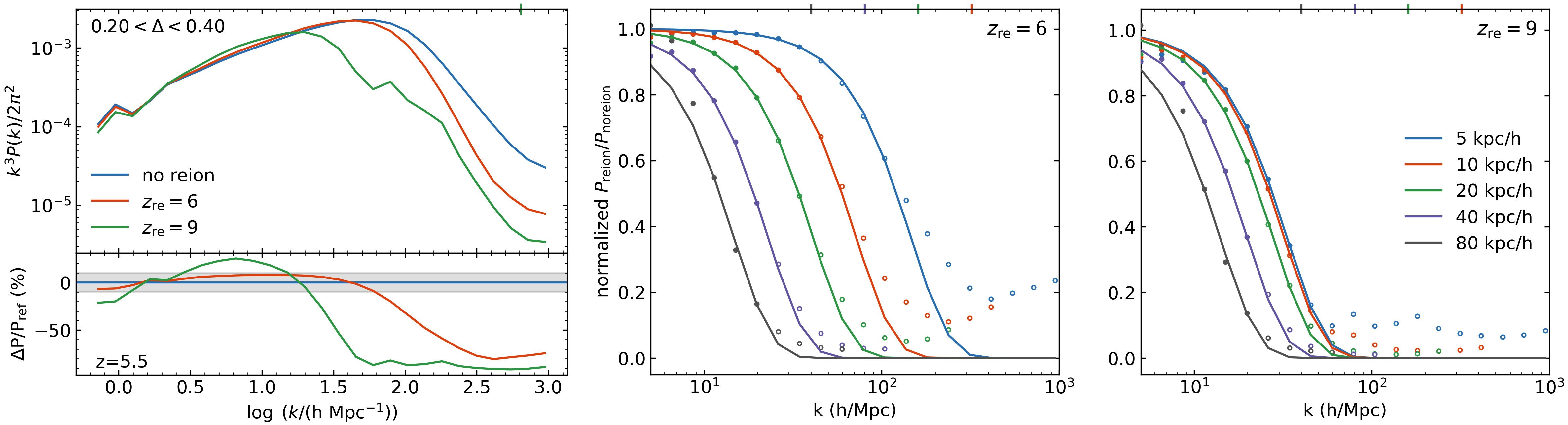}
\caption{\emph{Left panel:} The 3D power in baryons for 10 \mpch, 5 \kpch$\;$resolution simulations with the three reionization histories, with an overdensity clip $0.2 < \Delta < 0.4$ applied to highlight variation at relevant overdensities for the Ly$\alpha$ forest. Using the raw density fields, the power is $\sim$5 per cent higher for the $z_\mathrm{re}=6$ simulation than for the earlier flash, with a larger difference at high $k$. \emph{Right two panels:} The ratio of 3D gas power in reionized simulations with varying resolution to that of a high-resolution, unreionized one. All boxes are 10 \mpch$\;$boxes, and the reference unreionized simulation has $\Delta x=5$ \kpch. The colored points indicate the measured ratios from the simulations and the midpoint of each $k$ bin. Assuming that $P_\mathrm{reion}/P_\mathrm{noreion}$ goes as Equation~\ref{eq:puchwein22}, we fit to the solid points for $N$ and $k_\mathrm{ps}$ to produce the models, shown as the solid lines. Empty points are not included in the calculation, to avoid oscillatory trends in power at high $k$~\citep{gnedin98,puchwein22}, and also to avoid overfitting on highly sampled $k$ bins.}
\label{fig:pressure_smoothing}
\end{figure*}

Having visually established the effect of a late reionization pressure smoothing and the impact on Ly$\alpha$ power, it is of interest to quantitatively define the pressure smoothing scales in our two flash reionization scenarios. We plot the 3D power for the 10 \mpch, $\Delta x=5$ \kpch$\;$ boxes in the left panel of Figure~\ref{fig:pressure_smoothing} for the three reionization scenarios. \footnote{While not shown here, we find that for the raw 3D power the curves are practically identical; evidently the majority of the power at all physical scales is unaffected by the reionization model used in \texttt{Nyx}.} In order to select the underdense gas that gives rise to the Ly$\alpha$ forest, we apply an overdensity clip $0.2 < \Delta < 0.4$, equal to the range used for analysis of the pressure smoothing scale in~\citet{puchwein22}. We show these overdensity-restricted $P_\mathrm{3D}$ for the three reionization histories, with residuals calculated with respect to the \emph{no reion.} case. With this adjustment applied, the curves begin to differ substantially, with an earlier reionization showing greater power at $\log (k/(h\;$Mpc$^{-1}))\approx 0.8$ 
and reduced power at both smaller and larger wavenumbers, though the dropoff is much greater for larger $k$~\citep{kulkarni15,onorbe19}.

To quantify this dropoff, we replicate the method described in~\citet{puchwein22}, comparing the dimensionless 3D power in baryons in each reionized 10 \mpch$\;$simulation to the corresponding \emph{no reion.} box. Effectively this is similar to comparing the power in baryons to the dark matter~\citep[see e.g. Equation 3 in][]{lukic15}, although it is not completely identical as the reference \emph{no reion.} boxes still contain shock heating from structure formation at $z<10$ that will affect the distribution of baryons.

The power in the reionized simulation is assumed to follow an exponential cutoff of form
\begin{equation}\label{eq:puchwein22}
    P_\mathrm{re}^{\Delta x}(k) = N \cdot P_\mathrm{nore}^{5 \mathrm{kpc/h}}(k) \cdot \mathrm{exp} \left( -k^2 / k_\mathrm{ps}^2 \right)^2
\end{equation}
where $P_\mathrm{re}^{\Delta x}$ and $P_\mathrm{nore}^{5 \mathrm{kpc/h}}$ are the power spectra for the reionized and unreionized simulations, respectively, with the unreionized reference simulation always being $\Delta x=5$ \kpch. The value $N$ is a normalization constant and $k_\mathrm{ps}$ is the wavenumber corresponding to the pressure smoothing length $\lambda_\mathrm{ps}$, below which $P_\mathrm{3D}$ drops precipitously. To perform the fit, we maximize a multivariate Gaussian likelihood where, for the mean power $P(k)$ in each $k$ bin, the uncertainty is given by the standard error of the mean, or the standard deviation in power divided by the number of contributing $k$ modes. We assume that there is no correlation between the different $k$ modes.

The right two panels of Figure~\ref{fig:pressure_smoothing} show the ratio $P_\mathrm{re}^{\Delta x} / P_\mathrm{nore}^{5 \mathrm{kpc/h}}$, normalized by $1/N$, as points, while the model fits are shown as the solid lines. Evaluated by eye, the model does an acceptable job of describing the cutoffs, although the $\Delta x=80$ \kpch$\;$case is not steep enough to fit the measured power.\footnote{We find that we need to adjust the $k$ bins permitted in the fit to avoid overfitting on the high values, which are weighted heavily due to the number of $k$ modes contained therein.} The curves for varying resolutions in the early reionization case (rightmost panel) show convergence by $\Delta x = 10$ \kpch, while the later $z_\mathrm{re}=6$ history in the center panel does not converge even by $5$ \kpch. This visual assessment is supported by the fit results for $\lambda_\mathrm{ps}$ (see Table ~\ref{tab:fit_results}). For $z_\mathrm{re}=9$ the rate of change with $\Delta x$ slows by 10 \kpch, with $\lambda_\mathrm{ps}$ asymptoting to $\sim133$ \kpch. For $z_\mathrm{re}=6$, even for the change from 10 to 5 \kpch, the fit pressure smoothing scale still drops from 65 to 30 \kpch. This highlights the inability of the low resolution simulations to produce adequate $P_\mathrm{re}^{\Delta x}$ for larger wavenumbers.

An alternative estimate of the expected smoothing scale could be given by the filtering scale described in~\citet{gnedin98}, which accounts for the thermal history of the gas, including its current temperature. In a simplified form for high redshifts~\citep{puchwein22}, this is calculated as
\begin{equation}\label{eq:filtering_scale}
\lambda_F^2 = \frac{2}{a H_0^2 \Omega_m} \int_0^a da' a'^{\frac{3}{2}} \left[\frac{1}{\sqrt{a'}} - \frac{1}{\sqrt{a}} \right]c_s^2 \left(a'\right)
\end{equation}
with the sound speed a function of the scale factor $a$. For our highest resolution simulations, the anticipated filtering scales (calculated using Equation~\ref{eq:filtering_scale}, with $\lambda_\mathrm{f}=1/k_\mathrm{f}$) are 11 and 39 \kpch$\;$for the late and early reionization scenarios, or $\sim$2 times the quantities measured via the fits from Equation~\ref{eq:puchwein22} (see columns 4 and 7 in Table~\ref{tab:fit_results}).

While the low resolution simulations do not accurately reproduce the power $P_\mathrm{re}^{\Delta x}$, we find that they can still be used to measure an accurate value for the pressure smoothing scale calculated from the high resolution simulations. To accomplish this, the $P_\mathrm{nore}$ term in Equation~\ref{eq:puchwein22} simply needs to be taken from the same resolution simulation as $P_\mathrm{re}$. The fitted $\lambda_\mathrm{ps}$ increases slightly with worsening resolution, but only up to $\sim$ 36 \kpch$\;$ by 80 \kpch$\;$for $z_\mathrm{re}=6$. Thus, the pressure smoothing scale measured in this fashion can be determined even when the simulation resolution itself is worse than $\lambda_\mathrm{ps}$.

\subsection{Mean flux convergence}\label{ssec:mean_flux_convergence}
\begin{figure}
\hspace*{-0.1in}
\includegraphics[width=0.5\textwidth]{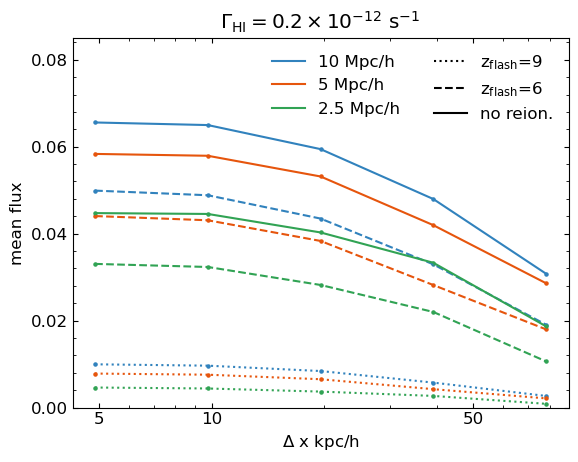}
\caption{Mean flux in Ly$\alpha$ transmission power spectra at $z=5.5$ for a constant UVB, as a function of the spatial resolution. For the given $\Gamma_\mathrm{HI}$ (required to match the mean flux in our largest, highest resolution simulation) and reionization history, a larger simulation box results in a higher mean flux given the same spatial resolution; a later reionization gives the same qualitative result.}\label{fig:mean_flux_Deltax}
\end{figure}

A consistent step in this analysis has been the practice of mean flux matching, rescaling the optical depth in each set of skewers so that the Ly$\alpha$ flux values amount to a reasonable average. But the mean flux itself can serve as a metric of convergence when the photoionization rate is a given; we investigate this as a function of the spatial resolution, box size, and reionization history in Figure~\ref{fig:mean_flux_Deltax}. We use a constant photoionization rate of $\Gamma_\mathrm{HI}=0.2 \times 10^{-12}$ s$^{-1}$, which is required for the 10 \mpch, 5 \kpch$\;$ simulation to match $\langle F\rangle=0.05$ used in the remainder of this work. Using this value, we find that for a constant box size and reionization history, the mean flux will be higher at a higher resolution, and will approximately converge for $\Delta x=10$ \kpch. For a given spatial resolution, a larger box size results in a higher mean flux; for example, at the highest resolution for $z_\mathrm{re}=6$ the mean flux in a $L_\mathrm{box}=2.5$ \mpch$\;$ is about 70 per cent that of a 5 \mpch$\;$ box. This confirms previous results from ~\citep[e.g.][]{lukic15}. Finally, a later reionization also results in a higher mean flux for a given box size and given resolution, likely a side effect of the higher average temperature, which increases the range of overdensities that contribute to high transmission cells.

For constant $\Gamma_\mathrm{HI}$, higher resolution simulations and larger box sizes lead to a higher mean flux. This arises from these simulations increased ability to model low overdensities, as seen in Figure~\ref{fig:pdf_resolution_boxsize}. Thus, the overall convergence of Ly$\alpha$ properties is related to a simulation's ability to produce an adequate number of low density, highly transmissive cells.

\begin{table}
  \centering
  \tablewidth{\textwidth}
  \caption{Results of fit to Equation~\ref{eq:puchwein22}, with $\lambda_\mathrm{ps}=2\pi/k_\mathrm{ps}$. In columns 4 and 7 we also explicitly give $1/k_\mathrm{ps}$ to adhere to the convention used for the filtering scale, which excludes the factor of $2\pi$.}\label{tab:fit_results}
  \begin{tabular}{ c c c c c c c }
    \hline
     & \multicolumn{3}{c}{$z_\mathrm{re}=6$} &  \multicolumn{3}{c}{$z_\mathrm{re}=9$} \\
     \makecell{$\Delta x$\\(\kpch)} & $N$ & \makecell{$\lambda_\mathrm{ps}$\\(\kpch)} & $1/k_\mathrm{ps}$ & $N$ & \makecell{$\lambda_\mathrm{ps}$\\(\kpch)} & $1/k_\mathrm{ps}$ \\
     \hline
     \\[-0.08in]
     39.0 & 1.77 & 198.2 & 31.6 & 1.82 & 226.7 & 36.1 \\
     19.5 & 1.55 & 115.0 & 18.3 & 1.74 & 163.8 & 26.1 \\
     9.8 & 1.31 & 65.3 & 10.4 & 1.68 & 139.1 & 22.1 \\
     4.9 & 1.13 & 30.5 & 4.9 & 1.66 & 133.6 & 21.3 \\
    \hline
  \end{tabular}
\end{table}

\section{Discussion}\label{sec:discussion} 
\subsection{Comparison with previous studies}\label{ssec:comparison_to_old_studies}
Previous investigations have shown that the mean flux values obtained in Ly$\alpha$ power spectra are affected by the resolution and volume of the simulations~\citep{tytler09,lukic15,chabanier22}. We confirm that smaller boxes result in smaller mean fluxes when subject to the same UVB and these do not converge by box sizes of $L_\mathrm{box}\sim10$ \mpch. On the other hand, $\langle F \rangle$ \emph{generally} increases for decreasing spatial resolutions until converging, all else being equal. We find that the impact of low resolution, in terms of phase space sampling, is quite similar to that of a small box, perhaps suggesting that some difficulties associated with small boxes can be alleviated with sufficient resolution. \citet{tytler09} finds that certain $z=2$ Ly$\alpha$ metrics that are not converged in small boxes can be ``corrected'' to match those found in larger boxes by injecting additional heat, which is naturally present in larger simulations.

We do note some small deviations in the observed rates of convergence compared to other studies. \citet{boera19} quantified convergence with resolution and box size when implementing a scaled version of the UVB from~\citet{hm12} starting at $z=15$, which itself leads to an early reionization and associated heating~\citep{puchwein15,onorbe17_2}. As in our result, they found that an increase in spatial resolution resulted in greater power for $\log (k/(\mathrm{s\;km}^{-1})>-1.6$ at $z=4.2$, 4.6, and 5.0 snapshots. The residual increased with redshift, exceeding 25 per cent for $\log (k/(\mathrm{s\;km}^{-1})>-1.0$ for an average spatial resolution of 40 \kpch$\;$compared to a 13 \kpch$\;$reference simulation, which is quantitatively similar to our results at $z=5.5$ with $z_\mathrm{re}=6$. Their results with varying box size differ more from ours, as while both studies show that larger boxes result in less small scale power, they find a residual that increases with $k$ for $\log (k/(\mathrm{s\;km}^{-1})>-1.0$ between $L_\mathrm{box}=10$ and 40 \mpch$\;$boxes, whereas our results show much smaller variation for smaller boxes.

The investigation in~\citet{chabanier22} similarly uses \texttt{Nyx}, with $L_\mathrm{box}=10$ \mpch, and finds varying rates of convergence. In particular, for $P_\mathrm{Ly\alpha}$ of $\log (k/(\mathrm{s\;km}^{-1}))\approx-1$ they find only a 7  per cent difference between 20 and 10 \kpch$\;$resolutions, whereas we find $\sim20$ per cent in our $z_\mathrm{re}=6$ model, and a similar quantity in our earlier flash. However, it is expected that the differences will be greater, as the LAF is known to be less converged closer to reionization~\citep{bolton09,lukic15,chabanier22}, and our comparison is at $z=5.5$ versus theirs at $z=5$, thus maintaining consistency with that expectation.

\citet{wu19} explored the impact of thermal fluctuations on the Ly$\alpha$ forest in {\sc arepo-rt} simulations, including variable inhomogeneous and flash reionization histories. As we describe in subsection~\ref{ssec:reionization_history}, and as is expected from convention, they find that there is some cancellation of the effects of reduced pressure smoothing by thermal broadening that occurs for reionizations with different timings. They also find that thermal fluctuations associated with an inhomogeneous reionization process introduce a reduction in power at $\log (k/(\mathrm{s\;km}^{-1}))>-1.0$ of 10 per cent. Thus, our use of a flash reionization may amplify power at small scales and enhance differences between resolutions.

The effect of patchy reionization on pressure smoothing is examined in~\citet{puchwein22}, where the inhomogeneous topology is found to result in losses in small scale power in reionized regions. In that work, which originates the method of measuring pressure smoothing used in Section~\ref{ssec:measuring_pss}, power spectra within the simulation are calculated for subvolumes, and the smoothing scale is determined as a function of the time since the subvolume reionized, $\Delta t_\mathrm{re}$. For a point of comparison, our $z_\mathrm{re}=9$ model reionizes $\sim$ 500 Myr prior to the snapshot containing our skewers, and produces $2 \pi/k_\mathrm{ps}=134$ \kpch, quite similar to their result ($\approx$129 \kpch, extracted from their Figure 13). The $z_\mathrm{re}=6$ model reionizes too recently to compare with their direct results from patchy reionization, as the smallest $\Delta t_\mathrm{re}$ is 200 Myr with a fit $2 \pi/k_\mathrm{ps}\approx$54 \kpch. Linearly interpolating between results from our two flash scenarios, we similarly arrive at $2\pi/k_\mathrm{ps}=55$ \kpch. Their free expansion model, in which the comoving smoothing length varies linearly with time, approximately matches the $\approx 30$ \kpch$\;$scale we arrive at for our $z_\mathrm{re}=6$, 5 \kpch$\;$resolution simulation at $\Delta t_\mathrm{re}=110$ Myr. Though not shown here, by re-running the two flash scenarios with different $\Delta T_\mathrm{re}$, we find that the heat added at reionization changes the measured $k_\mathrm{ps}$, with a smaller $\Delta T_\mathrm{re}$ producing a smaller smoothing scale for the same $\Delta t_\mathrm{re}$.

\subsection{Caveats}
In accordance with the $\Lambda$CDM model this investigation has assumed that dark matter is cold, but the introduction of thermal relic warm or fuzzy dark matter models could certainly reduce the resolution requirements for accurate IGM modeling, with the degree of reduction affected by the particle's characteristic streaming scale. Studies of Milky Way substructure place constraints on $m_\mathrm{WDM}$ to be $\gtrapprox2$ keV~\citep{polisensky11,newton21,dekker22}, while gravitational lensing studies and higher redshift results raise limits to higher masses, which produce even less of an impact on the small scale power~\citep{viel13b,gilman20}.~\citet{viel13b}, using Ly$\alpha$ forest observations, constrains models to those which deviate from CDM by less than 10 per cent at $k=10$ $h$ Mpc$^{-1}$ ($\log (k/\mathrm{(s\;km^{-1}}))=-0.83$ at $z=5.5$). Therefore, while WDM models are still viable, their masses appear to be high enough that they do not substantially diminish the small scale power with respect to the CDM prediction, and the need for a thorough understanding of the small scale convergence remains.

X-ray heating in some form is guaranteed as X-rays are generated by binary stellar remnants, particularly high mass X-ray binaries~\citep{gilfanov04,fabbiano06,mineo12}. Their long mean free paths would result in heat deposition directly into the IGM, assuming they could effectively escape the dense gas in their sourcing galaxies~\citep{madau97,das17}. While this is not strongly constrained, recent 21 cm analyses may imply a non-negligible degree of X-ray heating, placing the 95 per cent constraints on the gas kinetic temperature of $13 < T < 4768$ K at $z=7.9$~\citep{hera22a,hera22b}. If this heating occurs early enough in time, leaving enough time for the IGM to thermodynamically respond, then this would reduce the resolution necessary to accurately model the small scale power.

For simplicity, we have used a flash-type reionization, a limited number of reionization timings, and a constant amount of heat injection. In reality, hydrogen reionization is certainly an inhomogeneous and protracted process, occurring earlier in overdense regions with higher quantities of ionizing photons. While it will realistically start quite early, when the first generations of galaxies formed and began producing ionizing photons, it could not have progressed significantly until the ionizing photon budget was sufficiently large to outpace recombination. On the whole, this would result in substantial spatial variation in the effective smoothing scale, and by extension the overdensity, temperature, and velocity fields. Given that we recover the smoothing scales found in the patchy simulation subvolumes in~\citet{puchwein22}, however, it seems reasonable that accounting for inhomogeneity would result in a $P_\mathrm{Ly\alpha}$ that would be some composite of the cases examined here. For example,~\citet{onorbe19} found that flash models produce LAF power spectra similar to inhomogeneous models wherein the reionization midpoint matches $z_\mathrm{re}$.

\section{Conclusions}\label{sec:conclusions}
We have studied the convergence of the 3D and Ly$\alpha$ power spectra in \texttt{Nyx} cosmological simulations, focusing in particular on the effects of resolution, box size, and reionization history. With regards to resolution, we find that for a model that flash reionizes at $z_\mathrm{re}=6$, the 3D power converges to within 10 per cent for $\log (k/(h$ Mpc$^{-1}))<1.6$ by $\Delta x = 10$ \kpch, and that convergence in the dark matter power is better than for gas. $P_\mathrm{Ly\alpha}$ is converged up to $\log (k/($s km$^{-1}))=-1.0$ for $\Delta x=10$ \kpch, but for larger wavenumbers is too low by $10-20$ per cent with respect to our highest resolution simulation of 5 \kpch. An investigation of the PDFs and temperature-density phase space diagrams reveals that a higher resolution naturally leads to a wider dynamic range in probed values of temperature, peculiar velocities, and overdensities, so the wings of these distributions are quite populated for $\Delta x=5$ \kpch, and not for e.g. 80 \kpch. Higher resolutions lead to a higher mean flux for a given photoionization rate, and the mean flux convergence for a 10 \mpch$\;$box occurs at $\Delta x\sim10$ \kpch. The broad strokes of these results are in agreement with those of previous studies.

Turning to the box size, the 3D power is not converged by a $L_\mathrm{box}=10$ \mpch$\;$with $\Delta x=5$ \kpch, with the overall power in both gas and dark matter still increasing with $L_\mathrm{box}$, and the power in gas is less converged than that in dark matter. The Ly$\alpha$ power on scales approaching the mode of the box lies too low for boxes smaller than 10 \mpch$\;$with respect to that with $L_\mathrm{box}=10$ \mpch, although power for $\log (k/$(s km$^{-1}$))$>-1.5$ scales is converged within approximately 10 per cent by 5 \mpch. Power in a very small 1.25 \mpch $\;$box lies significantly too high, matching results from previous studies. The mean flux convergence is significantly affected by the box size, with $\langle F \rangle$ for a given $\Gamma_\mathrm{HI}$ increasing with box size. Larger boxes, sample a wider range of physical properties than smaller boxes. This naturally leads to greater variation around the mean flux, and an increase in power on small scales.

We have also varied the fiducial assumption of reionization occurring at $z=6$, and accounted for very early and very late models. Implementing an early ($z=9$) reionization causes a reduction in variance on small scales, and conversely an unreionized model permits survival of this variance to later times. This is due primarily to the stronger small scale fluctuations in overdensity, with gains in the inner regions of filaments and losses in their more extended outskirts; however, the visible degree of this effect in $P_\mathrm{Ly\alpha}$ is also dependent on the temperatures and velocities encountered by the skewers. For example, even gas in an un-reionized IGM will have less large $k$ Ly$\alpha$ power than a reionized one if the heat injection is large enough. The mean flux is affected by the reionization history, with the same resolution and box size having a substantially higher mean flux with a later reionization timing while holding the H I photoionization constant. This is due to the higher temperature permitting greater Ly$\alpha$ transmission in denser gas, similar in impact to varying the photoionization rate.

The physical scale at which the $P_\mathrm{3D} (0.2 < \Delta < 0.4)$ cuts off in low resolution simulations responds strongly to the reionization timing. For $z_\mathrm{re}=6$, we measure $\lambda_\mathrm{ps}=30$ \kpch$\;$where for $z_\mathrm{re}=9$ it is 133 \kpch$\;$. For the later flash simulation this scale is not evidently converged down to our highest resolution of $\Delta x=5$ \kpch, with the measured value drastically changing with increasing resolution, while for an early reionization it converges by $\Delta x=10$ \kpch. However, we also find that low resolution simulations can still reproduce the correct ratio of reionized to unreionized power, and $\lambda_\mathrm{ps}$ is recoverable using simulations with resolution down to 40 or even 80 \kpch. The measured pressure smoothing scales almost perfectly match those inferred in~\citep{puchwein22}, and adhere to the expectations of their simple expansion model framework. We do find that a smaller heat injection ($<20000$ K) at reionization causes a reduced pressure smoothing scale to be recovered using these methods.

In summary, we find that for a simulation using 20 \kpch$\;$ resolution, the Ly$\alpha$ forest power spectrum at $z=5.5$ is not converged within 10 per cent for $\log (k/($s km$^{-1}))>-1.4$, lying 30 per cent lower than a 5 \kpch$\;$ model at $\log (k/($s km$^{-1}))=-1.0$. Resolutions on the order of 10 \kpch$\;$ are necessary for a per cent level convergence up to $\log (k/($s km$^{-1}))<-1.1$, but for higher $k$ this may still be insufficient. Further, while this requirement is slightly more stringent for later reionizations, it does not differ significantly, likely as a result of the increased thermal broadening that occurs when reionization is more recent. The focus of simulations of the high-$z$ LAF should therefore be on achieving properties that adequately model the underdensities significantly impacting the Ly$\alpha$ mean flux and the variance in the flux contrast field. We suggest resolutions of 10 \kpch$\;$ or smaller and boxes larger than 10 \mpch$\;$ are necessary to accomplish this.

\section*{Acknowledgements}
CD acknowledges helpful conversations with the ENIGMA group at UC Santa Barbara and Leiden University. This research used resources of the National Energy Research Scientific Computing Center, which is supported by the Office of Science of the U.S. Department of Energy under Contract No. DE-AC02-05CH11231. ZL acknowledges the support from the Exascale Computing Project (17-SC-20-SC), a collaborative effort of the U.S. Department of Energy Office of Science and the National Nuclear Security Administration.  JO acknowledges the support from Ministerio de Ciencia, Innovación y Universidades (Spain) (“Beatriz Galindo” Fellowship BEAGAL18/00057).

\section*{Data Availability}
The data will be shared on reasonable request to the corresponding author.



\bibliographystyle{mnras}
\bibliography{mnras_template} 



\appendix
\section{Error advection in \texttt{Nyx}} \label{sec:appendixa}
\begin{figure}
\hspace*{-0.1in}
\includegraphics[width=0.5\textwidth]{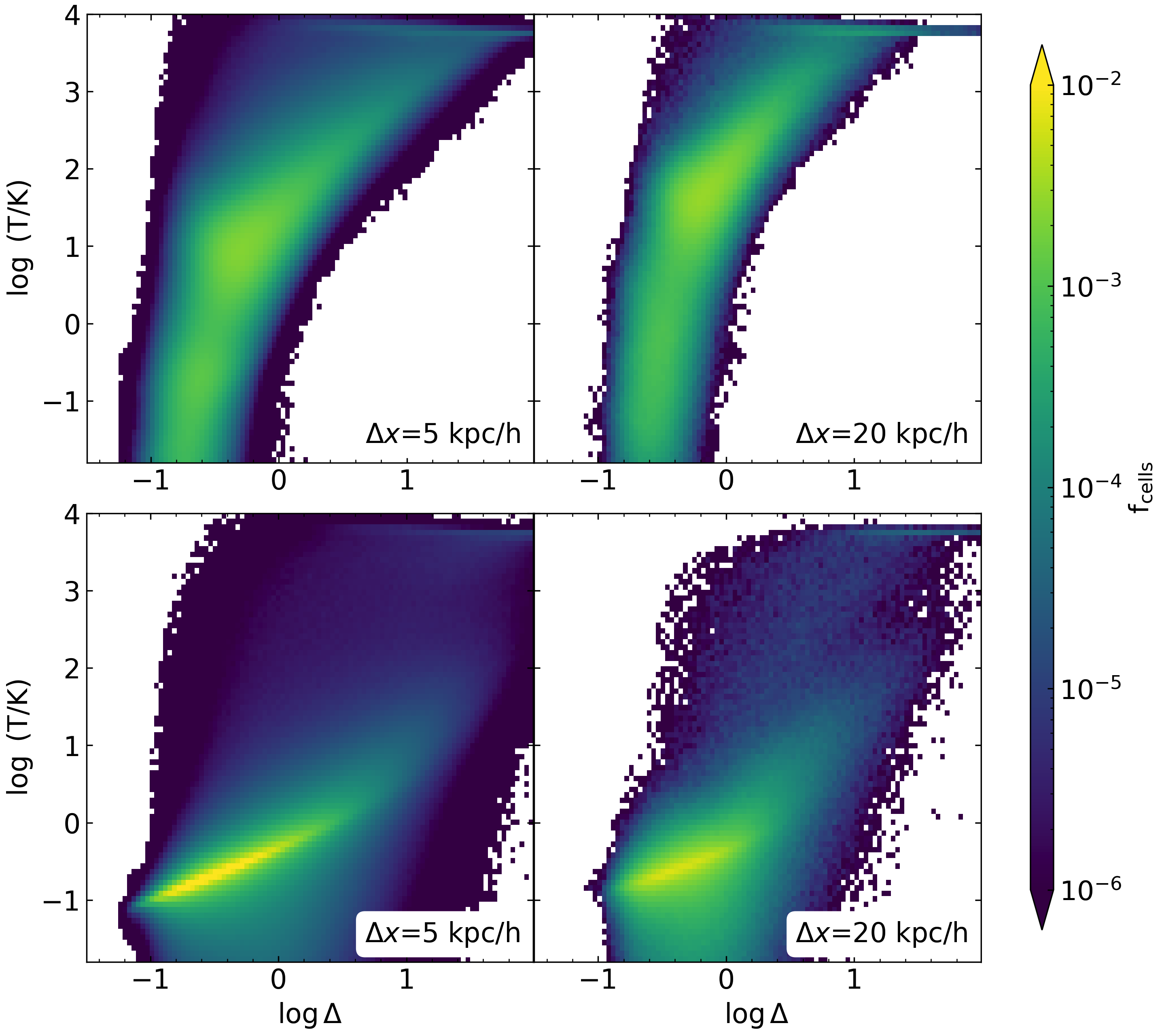}
\caption{\emph{Top row:} Temperature-density phase space for an unreionized 2.5 \mpch$\;$\texttt{Nyx} simulation at $z=5.5$ with the default treatment for internal energy, for two resolutions. For an unmodified treatment, the cell distribution is unnaturally bifurcated and does not adhere to the expected adiabatic slope. However, the distortion may begin to improve with increasing resolution. \emph{Bottom row:} A modified treatment where the transfer of energy is modulated in regions with extreme temperature gradients, with $\eta=0.1$ (see Equation~\ref{eq:eta_energy_treatment}). Even for the lower resolution considered here of 20 \kpch, the distribution looks approximately correct.}
\label{fig:rhot_eta_resolution}
\end{figure}

\begin{figure}
\hspace*{-0.1in}
\includegraphics[width=0.5\textwidth]{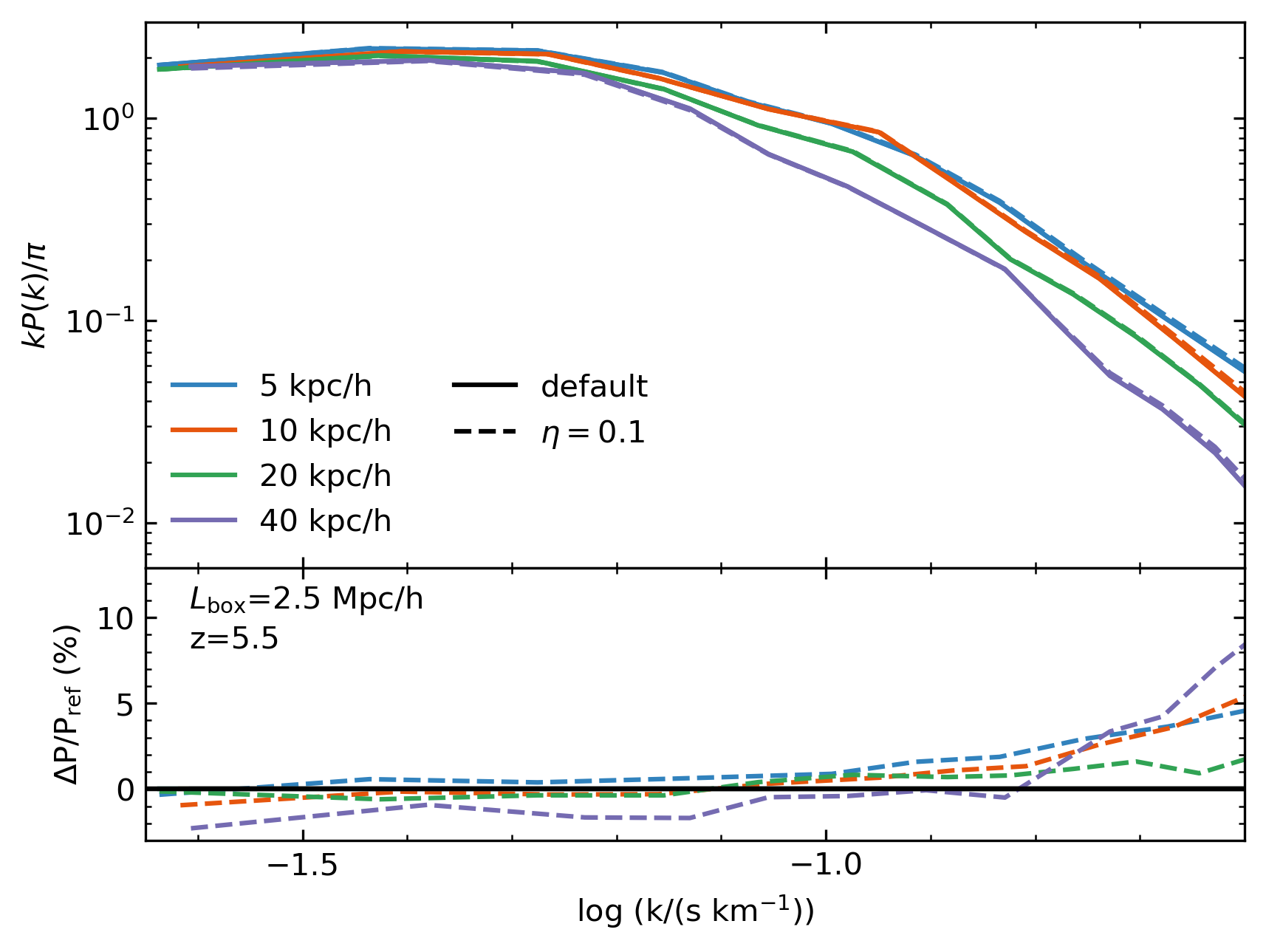}
\caption{The 1D power in Ly$\alpha$ for the simulations in Figure~\ref{fig:rhot_eta_resolution}, with additional resolution $\Delta x=10$ and 40 \kpch, with the unmodified internal energy treatment (solid) alongside the model with $\eta=0.1$ (dashed). The residuals for each resolution are plotted in the bottom panel with respect to the \emph{default} model. The residuals are all within 10 percent for this range of wavenumbers.}
\label{fig:PLya_eta_comparison}
\end{figure}

Due to numerical effects, it is possible for codes relying on the piecewise parabolic method to generate large errors in the internal energy $e$ when the total energy $E$ is dominated by a kinetic component $\rho \vec{v}^2/2$, for example in shocked regions with high Mach numbers. These errors in the total energy are then advected to neighboring cells, causing an accumulation of uncertainty in the temperature distribution, although the dynamics of the gas are unaffected. In practice this results in spurious heating and an unphysical vertical bifurcation in the temperature-overdensity plane prior to reionization (Figure~\ref{fig:rhot_eta_resolution}).
Codes such as {\sc Enzo} apply a corrective method described in~\citet{bryan95}, which compares the value of the internal energy to the total in the adjacent cells
\begin{equation}\label{eq:eta_energy_treatment}
    e = 
    \begin{cases}
    \left(E - \vec{v}^2/2\right), & \rho \left( E-\vec{v}^2/2\right) / \mathrm{max\;adjacent\;}(\rho E)> \eta\\
    e, & \rho \left( E-\vec{v}^2/2\right) / \mathrm{max\;adjacent\;}(\rho E) < \eta
    \end{cases}
\end{equation}
Where the threshold $\eta$ is exceeded, the internal energy is redefined as the value expected from the equation of state; when not, the independently calculated value is left as-is. Figure~\ref{fig:rhot_eta_resolution} shows the $\rho-T$ distribution for an unreionized 2.5 \mpch$\;$box at $z=5.5$ for two spatial resolutions with \emph{no correction} for the internal energy (top row) and with $\eta=0.1$ (bottom row). The uncorrected distribution shows the erroneous bifurcation, and the corrected simulation results in a more reasonable distribution, with the fix having shifted the gas nearer to the adiabatic relation.

The Ly$\alpha$ power for these simulations (plus some additional resolutions) are shown in Figure~\ref{fig:PLya_eta_comparison}, where we find that the differences are, perhaps surprisingly, quite minor. While for 40 \kpch$\;$they differ up to 8 per cent at $\log (k/($s km$^{-1}))=-0.6$, for higher resolutions they differ by $\leq5$ per cent. The simulations including $\eta=0.1$ consistently increase the power for $\log (k/($s km$^{-1}))>-1.1$, and the increase grows with wavenumber, so the reduction in excess heating is preserving additional high $k$ structure. However, it is unclear if the changes are exaggerated as the implementation of Equation~\ref{eq:eta_energy_treatment} can also cause an oversuppression of shock heating if $\eta$ is too high.

Overall, these changes are relatively minor compared to the differences introduced by the other variables we have examined here. It may be that, once the heat is injected, all that matters for the scales considered here is whether the appropriate overdensity regime is sufficiently sampled.

\section{Impact of dark matter-baryon streaming velocities} \label{sec:appendixb}

\begin{figure*}
\includegraphics[width=1.0\textwidth]{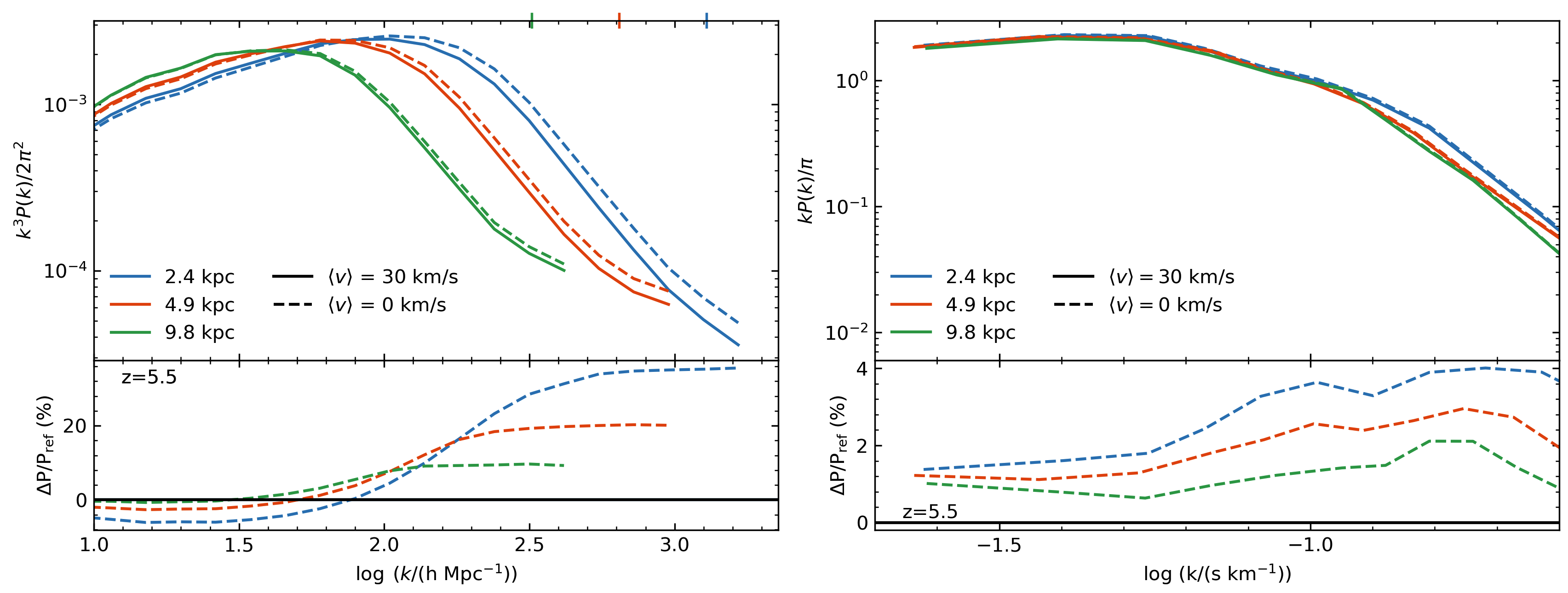}
\caption{Dimensionless power within $0.2 < \Delta < 0.4$ in the 3D baryonic matter distribution for an unreionized 2.5 \mpch$\;$box with varying spatial resolutions. We compare the power in the box using our fiducial initial condition setup, with an initial average baryon-dark matter streaming velocity of 30 km/s, to a setup with $v=0$ km/s. The simulations with $v=0$ km/s have increased power on scales $\log (k/($h Mpc$^{-1}))\sim1.8$ with respect to $v=30$ km/s. The residual between simulations with identical resolutions and different $\langle v \rangle$ (bottom panel) increases with decreasing spatial resolution, and is larger at large $k$. The corresponding Ly$\alpha$ power (right plot) reveals that the differences are quite small at the relevant scales, only 2-4 per cent higher for the zero streaming velocity. However, the residual does increase with increasing spatial resolution.}
\label{fig:P3d_vstream_comparison}
\end{figure*}

When generating our initial conditions, we have elected to use the code {\sc CICASS}~\citep{oleary12} in the event that the non-zero dark matter-baryon streaming velocity leads to early shock heating, which could result in subsequent removal of small scale structure and an increase in the pressure smoothing. Figure~\ref{fig:P3d_vstream_comparison} shows the effect of streaming velocity in the initial conditions on baryonic power at $z=5.5$ for an unreionized box. As expected, the nonzero streaming velocity in the $\langle v\rangle =30$ km/s leads to a reduction in small scale power as seen in~\citet{oleary12}. This could be due to a combination of the physical motion of the particles, or a consequence of increased shock heating at early times.

With $\Delta T_\mathrm{re}=20000$ K, the Ly$\alpha$ power for these two $\langle v \rangle$ cases do not show much of a difference. with the higher resolution simulations showing higher power only by a maximum of about 4 percent. It may be that for a lower heat injection, the discrepancy would be larger. However, given that this is a reasonable temperature to use~\citep{miraldaescude94,mcquinn12,daloisio19}, it seems that the contribution to the $-1.0 < \log (k/(\mathrm{s \; km}^{-1})) < -0.6$ Ly$\alpha$ power from a nonzero streaming velocity is negligible compared to resolution effects.

\label{sec:appendixc}


\bsp	
\label{lastpage}
\end{document}